\documentclass[10pt,english]{article}
\usepackage[T1]{fontenc}
\usepackage[latin9]{inputenc}
\usepackage{geometry}
\usepackage{amsmath}
\usepackage{amssymb}
\usepackage{graphicx}
\usepackage{esint}
\PassOptionsToPackage{normalem}{ulem}
\usepackage{ulem}

\makeatletter

\providecommand{\tabularnewline}{\\}

\makeatother

\usepackage{babel}
\begin{document}

\title{EPRL/FK Asymptotics and the Flatness Problem}

\author{José Ricardo Oliveira%
\thanks{josercamoes@gmail.com%
}\\
\emph{\normalsize{School of Mathematical Sciences, University of Nottingham}}%
\thanks{Paper written as a result of PhD work under the supervision of Prof.
John W. Barrett. The author acknowledges funding by the Fundação para
a Ciência e Tecnologia (FCT), scholarship SFRH/BD/77045/2011.%
}\\
\emph{\normalsize{University Park}}\\
\emph{\normalsize{Nottingham NG7 2RD, UK}}}
\maketitle
\begin{abstract}
Spin foam models are an approach to quantum gravity based on the concept
of sum over states, which aims to describe quantum spacetime dynamics
in a way that its parent framework, loop quantum gravity, has not
as of yet succeeded. Since these models' relation to classical Einstein
gravity is not explicit, an important test of their viabilitiy is
the study of asymptotics - the classical theory should be obtained
in a limit where quantum effects are negligible, taken to be the limit
of large triangle areas in a triangulated manifold with boundary.
In this paper we will briefly introduce the EPRL/FK spin foam model
and known results about its asymptotics, proceeding then to describe
a practical computation of spin foam and semiclassical geometric data
for a simple triangulation with only one interior triangle. The results
are used to comment on the \textquotedbl{}flatness problem\textquotedbl{}
- a hypothesis raised by Bonzom (2009) suggesting that EPRL/FK's classical
limit only describes flat geometries in vacuum.
\end{abstract}

\date{\tableofcontents{}}

\section{Introduction}

Spin foam models are an approach to quantum gravity heavily inspired
in Loop Quantum Gravity (LQG)\cite{key-72}, which aimed to address
the parent theory's issues with describing dynamics while providing
a clear picture of the quantum geometry of a general relativistic
spacetime. While LQG's proposed canonical quantization of the first
order version of Einstein's general relativity gives us a background-independent
model with mostly well understood kinematics, its dynamics is encoded
in a time evolution equation named the Hamiltonian constraint - time
being defined via a 3+1 ADM decomposition of spacetime\cite{key-73}
used to derive a Hamiltonian form for the Holst-Palatini action. Until
today, solving the Hamiltonian constraint remains an open problem,
due to two main issues. One is the \emph{problem of time }- defining
the dynamics of a system which is manifestly diffeomorphism-invariant
by its time evolution is possible in the classical theory, by considering
the time variable in an ADM decomposition of a solution of Einstein's
equations, but in a quantum version of the theory these solutions
only determine probabilities of different spacetimes occurring, and
therefore defining time in this way would be ambiguous. While the
former is a more conceptual problem with known workarounds\cite{key-86},
there is also a more serious practical issue - quantizing the Hamiltonian
constraint and writing down the respective operator. There are several
ambiguities in doing so, and while there are proposals for it, such
as Thiemann's\cite{key-87}, it remains as an open problem, especially
because it has proven difficult to verify the viability of a given
proposal.\\
\\
The spin foam approach originated from an attempt to enunciate a path
integral formulation of LQG. It uses the basis of spin network states,
taking them as quantum states of a triangulated manifold which are
summed over to form a partition function. Dynamics is determined by
the probability amplitudes attributed to each state. Therefore, the
problem to solve in the spin foam program is to define a set of amplitudes
which is consistent with GR. Ponzano and Regge formulated a suitable
model for three-dimensional gravity\cite{key-74}, but the four-dimensional
problem is much more difficult in nature - 3d general relativity in
vaccuum is purely topological, with no dynamical degrees of freedom,
while the 4d theory is not\cite{key-75}.\\
\\
The first concrete attempt at devising a spin foam model for 4d gravity
was the Barrett-Crane model\cite{key-95}, which gave a set of bivector
variables, obtainable from the spin foam parameters, and equivalent
to a set of variables describing the Euclidean geometry of a triangulation.
The model was later abandoned as it was found that the bivectors were
over-constrained by the requirement of simplicity. The idea of enforcing
that specific constraint only in a weak ``expectation value'' sense
instead of the strong sense led to two independent proposals (Engle/Pereira/Rovelli/Livine
and Freidel/Krasnov), which turned out to be equivalent for Immirzi
parameter $0<\gamma<1$, and gave rise to the EPRL/FK model. Additionally,
the Ooguri\cite{key-76} and Crane-Yetter\cite{key-77} models are
often mentioned as triangulation-independent models that do not describe
gravity. \\
\\
Study of asymptotics becomes necessary for two main reasons. The first,
and most evident, is to determine if a given model reduces to classical
General Relativity in the $\hbar\rightarrow0$ limit. Secondly, it
seems apparent that diffeomorphism invariance in GR should be realized
as triangulation independence in the spin foam model, but this condition
is manifestly not satisfied in either the Barrett-Crane or EPRL/FK
models. While this is a major issue in itself, it can be argued that
the triangulation invariance requirement can be ``relaxed'' somewhat,
by only enforcing it in the semiclassical limit. The implication that
there would be ``preferred coordinate choices'', realized as preferred
triangulations, in the full quantum theory is certainly an uncomfortable
one, but not necessarily invalid, since the length scale which one
would have to probe to ``see'' the triangulation structure of spacetime,
if it exists, is well below anything feasible with the current means.\\
\\
In Section 2, we briefly introduce the basic concepts of general spin
foam models in four dimensions, and fully define the EPRL/FK model
in the Euclidean setting with an Immirzi parameter $0<\gamma<1$.
Section 3 is dedicated to the asymptotics of the EPRL/FK model in
an arbitrary simplicial complex with boundary, consisting of a short
review of past work and results, as well as more detailed considerations
about minute details in the formalism and the key tool used to derive
a semiclassical limit, the stationary phase method, leading into some
new insight on the ``flatness problem''.\\
\\
Finally, section 4 includes a thorough calculation of the classical
geometry of a simplicial complex dubbed $\Delta_{3}$, consisting
of three 4-simplices, describing the methods used which apply to any
Regge-like boundary data and presenting the results obtained from
two examples with specified boundary. Section 5 is reserved for discussion
of the results and their implications about the validity of the model.

\section{Spin Foam Models and EPRL/FK}

Spin foams are constructed from arbitrary spin network states $\psi_{\Gamma}\left(\left\{ g_{l}\right\} \right)$
over graphs $\Gamma$ embedded in a manifold $\mathcal{M}$ (which
corresponds to the spatial slice of spacetime), where $g_{l}$ are
elements of a gauge group $G$ which in gravity is the relativistic
symmetry group of the theory (in general it could be any Lie group).
The edges $l$ of $\Gamma$ have spins $j_{l}$ associated to them,
corresponding to irreducible representations of $G$, while the graph's
vertices $v$ are labelled by intertwiners $i_{v}$. Now if we picture
the extra time dimension and imagine the graph evolving into it, it
will form a so-called \emph{2-complex, }where the edges are foliated
into faces $f$ and the vertices into new edges $e$. The graph can
change topologically with time, and there will be new vertices $v$,
signalling points in spacetime where one edge breaks into several,
or vice-versa with two or more edges joining into one. The ``time-evolved''
graph is called the spin foam, and can be generally defined by
\begin{itemize}
\item an arbitrary 2-complex;
\item representation spins $j_{f}$ for each face $f$ of the 2-complex;
\item intertwiners $i_{e}$ for each edge $e$.
\end{itemize}
In four dimensions, the geometrical picture associated to spin foam
gravity can be described intuitively with the existent duality between
2-complexes and triangulations of a 4-dimensional manifold. Indeed,
a spin foam model in four dimensions can be defined as a state sum
whose quantum states are configurations of a $4$-dimensional simplicial
complex $\Delta$ with its $4$-simplices $\sigma_{v}$, tetrahedra
$\tau_{e}$ and triangles $\delta_{f}$ coloured by a set of geometrical
variables $c$ \cite{key-78}. $\Delta$ can be associated with its
dual 2-complex as follows:\\

\begin{center}
\begin{tabular}{|c|c|}
\hline 
simplicial complex & dual 2-complex\tabularnewline
\hline 
\hline 
$\sigma_{v}$ & vertex $v$\tabularnewline
\hline 
$\tau_{e}$ & edge $e$\tabularnewline
\hline 
$\delta_{f}$ & face $f$\tabularnewline
\hline 
\end{tabular}\\

\par\end{center}

\noindent The state sum is defined for a given simplicial complex,
and is a weighted sum over all possible colourings, with amplitudes
attributed to each face, edge and vertex. 
\begin{equation}
Z=\sum_{\text{colourings }c}\prod_{f}W_{f}(c)\prod_{e}W_{e}(c)\prod_{v}W_{v}(c)
\end{equation}
$W_{f},\, W_{e},\, W_{v}$ are the face, edge and vertex amplitudes
of each configuration, respectively. Defining a particular spin foam
model corresponds to setting these amplitudes. We now state them for
the EPRL/FK model\cite{key-79,key-80} in Euclidean signature.

\subsubsection*{Vertex amplitude $W_{v}$}

We follow the construction of $W_{v}$ given in\cite{key-84}. The
colourings for the Euclidean EPRL/FK model are SU(2) quantum numbers
$j_{f}$ for each face and SU(2) intertwiners $\hat{\iota}_{e}$ for
each edge, given by
\begin{equation}
\hat{\iota}_{e}\left(k_{ef},\, n_{ef}\right)=\int_{SU(2)}dh_{e}\bigotimes_{f\in e}h_{e}\left|k_{ef},\, n_{ef}\right\rangle 
\end{equation}
where $\left|k,\, n\right\rangle \equiv\left|k,\,\vec{n},\theta_{n}\right\rangle $
are the Livine-Speziale coherent states\cite{key-81} in the spin-$k$
representation of SU(2)%
\footnote{Note that a priori $k_{f}\neq j_{f}$.%
}. They minimize the uncertainty $\Delta(J^{2})=\left|\left\langle \vec{J}^{2}\right\rangle -\left\langle \vec{J}\right\rangle ^{2}\right|$
in the direction of angular momentum $\vec{n}$, and their definition
is
\begin{equation}
\left|k,\, n\right\rangle \equiv G(\vec{n})\left|k,\, k\right\rangle _{\vec{z}}
\end{equation}
where $\left|k,\, k\right\rangle _{\vec{z}}$ is the maximum angular
momentum eigenstate of $\hat{J}_{z}$ and $G(\vec{n})\in SU(2)$ rotates
$\vec{z}$ into $\vec{n}$. There is a phase ambiguity in this definition
that cannot be resolved in a canonical way, since the information
about it is lost in the projection of the state vector $\left|n\right\rangle \in S^{3}\subset\mathbb{C}^{2}$
to $S^{2}$ to obtain the rotation axis $\vec{n}$. It will become
apparent in a later section that this ambiguity is not reflected in
any calculations, as all related phase factors cancel out.\\
\\
For the intertwiner definition to make sense there must be an ordering
of the faces in a tetrahedron\cite{key-82}. Setting an ordering for
the points in a 4-simplex, $\sigma_{v}=(p_{1},p_{2},p_{3},p_{4},p_{5})\equiv(1,2,3,4,5)$,
is equivalent to doing the same for the tetrahedra in it, since the
tetrahedron $t_{e_{i}}$ can be defined as the one that does not contain
the point $i$. The operation
\begin{eqnarray}
\partial_{i}(v_{1},...,v_{n}) & \equiv & (-1)^{i}(v_{1},...,\hat{v_{i}},...,v_{n})\nonumber \\
\partial_{n+1}(v_{1},...,v_{n}) & \equiv & \partial_{n}(v_{1},...,v_{n})\label{eq:orient}
\end{eqnarray}
induces an ordering in a $(n-1)$-simplex from that of a $n$-simplex.
Using it, we can establish a coherent ordering of tetrahedra and triangles
starting from what was defined for the 4-simplex. We can also define
the orientation of a simplex - $(v_{1},...,v_{n})$ is positively
oriented if it is an even permutation of $(1,...,n)$, and negatively
oriented otherwise. Since $\partial$ satisfies $\partial_{i}\partial_{j}=-\partial_{j}\partial_{i}$,
a consequence of the definition is that if $f=t_{e_{1}}\cap t_{e_{2}}$,
then the orientations of $f$ induced by $t_{e_{1}}$ and $t_{e_{2}}$
are opposite. This has an intuitive explanation if one considers the
normal vectors to each tetrahedron. \\
\\
The construction of the 4-vertex amplitude is based on the spin network
basis states of Loop Quantum Gravity\cite{key-84}, and it relies
on defining a Spin(4) (that is, the Euclidean isometry group SO(4))
intertwiner $\iota_{e}$ from $\hat{\iota}_{e}$, using the decomposition
$\text{SU(2)}\times\text{SU(2)}=\text{Spin(4)}$. First note that
\begin{equation}
\hat{\iota}_{e}\in\text{Hom}_{\text{SU(2)}}\left(\mathbb{C},\,\bigotimes_{f\in e}V_{k_{ef}}\right),
\end{equation}
since it is a SU(2)-invariant vector of $\bigotimes_{f\in e}V_{k_{ef}}$,
where $V_{k_{ef}}$ is the vector space associated with the $k_{ef}$-spin
(irreducible unitary) representation of SU(2). One can construct an
injection
\begin{equation}
\phi:\,\text{Hom}_{\text{SU(2)}}\left(\mathbb{C},\,\bigotimes_{f\in e}V_{k_{ef}}\right)\rightarrow\text{Hom}_{\text{Spin(4)}}\left(\mathbb{C},\,\bigotimes_{f\in e}V_{j_{f}^{-},j_{f}^{+}}\right)
\end{equation}
such that $\phi(\hat{\iota}_{e})=\iota_{e}$ is the Spin(4) intertwiner.
This is done by using the Clebsch-Gordan maps $C_{k_{ef}}^{j_{f}^{-},j_{f}^{+}}:\, V_{k_{ef}}\rightarrow V{}_{j_{f}^{-}}\otimes V_{j_{f}^{+}}\approx V_{j_{f}^{-},j_{f}^{+}}$
and constraining the values of $j_{f}^{\pm}$ via the Immirzi parameter:
$j_{f}^{\pm}=\frac{1}{2}\left|1\pm\gamma\right|j_{f}$ relates them
to the original SU(2) quantum number (which is itself constrained
by this relation, since $j_{f}^{\pm}\in\frac{\mathbb{N}}{2}$). 
\begin{equation}
\iota_{e}(j_{f},\, n_{ef})\equiv\sum_{k_{ef}}\int_{\text{Spin(4)}}dg(\pi_{j_{f}^{-}}\otimes\pi_{j_{f}^{+}})(g)\circ\bigotimes_{f\in e}C_{k_{ef}}^{j_{f}^{-},j_{f}^{+}}\circ\hat{\iota}_{e}(k_{ef},\, n_{ef}),
\end{equation}
where $g=(g^{+},g^{-}),\, g^{\pm}\in\text{SU(2)}$ and $\pi_{j_{f}^{\pm}}:\,\text{Spin(4)}\rightarrow V_{j_{f}^{\pm}}$,
such that $(\pi_{j_{f}^{-}}\otimes\pi_{j_{f}^{+}})(g):\, V_{j_{f}^{-}}\otimes V_{j_{f}^{+}}\rightarrow V_{j_{f}^{-},j_{f}^{+}}$.
The integration over Spin(4) is there, once again, to ensure group
invariance of the intertwiner.%
\footnote{The sum over $k_{ef}$ is there because the edge amplitude has the
practical effect of selecting these numbers. For a general $W_{e}$,
they are summed over (as happens in the FK model for $\gamma>1$)%
}\\
The vertex amplitude $W_{v}$ is then a closed spin network (more
details on graphical calculus in\cite{key-89} for the Lorentzian
case) constructed by taking $\bigotimes_{e=1}^{5}\iota_{e}$ and ``joining
the extremities'', for each face, of the two edges that share it,
as illustrated in the figure below (each face corresponds to 2$\times$2
of the extremities, for a total of 40, since a 4-simplex has 10 tetrahedra)
by using the so-called $\epsilon$-inner product
\begin{equation}
\epsilon_{k}:V_{k}\otimes V_{k}\rightarrow\mathbb{C}.
\end{equation}

\begin{center}
\includegraphics[scale=0.95]{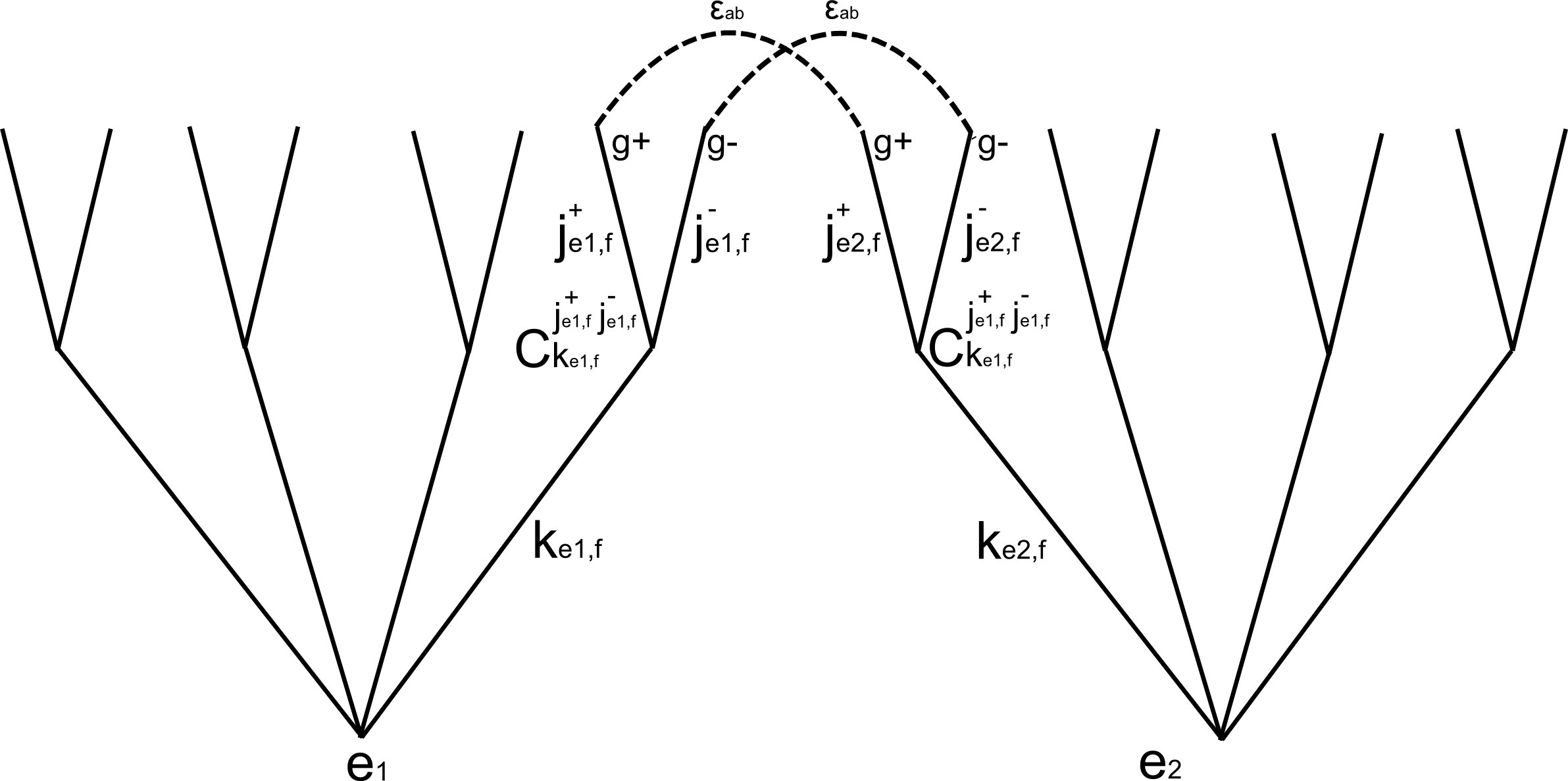}
\par\end{center}

\noindent The inner product is constructed by linearity from the $\epsilon_{1/2}$,
given in our convention by the matrix $\epsilon_{ab}=\left[\begin{array}{cc}
0 & i\\
-i & 0
\end{array}\right]$. The spin network diagram can now be evaluated using the Kaufmann
bracket\cite{key-88} with parameter $A=-1$. In practice this means
that each pair of crossing lines with spins $k_{1},\, k_{2}$ adds
a sign $(-1)^{4k_{1}k_{2}}$. These signs result in an overall sign
$(-1)^{\chi}$ in the amplitude.\\
\\
Finally, $W_{v}$ takes the form (now introducing the dependence in
$v$) 
\begin{equation}
W_{v}=(-1)^{\chi}\sum_{\{k_{ef}\}}\int_{\text{Spin(4)}^{5}}\prod_{e\in v}dg_{ve}^{+}dg_{ve}^{-}\int_{(S^{3})^{20}}\prod_{ef}dn_{ef}\left(\bigotimes_{f}\mathcal{K}_{vf}\right)\circ\left(\bigotimes_{e}\hat{\iota}_{e}\right)
\end{equation}
where{\small{
\begin{equation}
\mathcal{K}_{vf}=\left(\epsilon_{j_{f}^{-}}\otimes\epsilon_{j_{f}^{+}}\right)\circ\left[\left(\left(\pi_{j_{f}^{-}}(g_{ve_{f}}^{-})\otimes\pi_{j_{f}^{+}}(g_{ve_{f}}^{+})\right)\circ C_{k_{ef}}^{j_{f}^{-},j_{f}^{+}}\right)\otimes\left(\left(\pi_{j_{f}^{-}}(g_{ve'_{f}}^{-})\otimes\pi_{j_{f}^{+}}(g_{ve'_{f}}^{+})\right)\circ C_{k_{e'f}}^{j_{f}^{-},j_{f}^{+}}\right)\right].
\end{equation}
}}In this expression, $e,\, e'$ are the edges that share the face
$f$.

\subsubsection*{Edge amplitude $W_{e}$}

The edge amplitude is taken in modern models to be a selection rule
for the values of $k_{ef}$, and is the only difference between the
EPRL and FK models. Its choice depends on the value of the Immirzi
parameter.
\begin{itemize}
\item for $\gamma<1$, both EPRL and FK select the choice $k_{ef}=j_{f}=j_{f}^{+}+j_{f}^{-}$:
\begin{equation}
W_{e}^{\gamma<1}=d_{\hat{\iota}_{e}}\prod_{f\in e}\delta_{k_{ef},\, j_{f}^{+}+j_{f}^{-}}
\end{equation}

\item for $\gamma>1$, EPRL select $k_{ef}=j_{f}=j_{f}^{+}-j_{f}^{-}$,
\begin{equation}
W_{e}^{\text{EPRL},\,\gamma<1}=d_{\hat{\iota}_{e}}\prod_{f\in e}\delta_{k_{ef},\, j_{f}^{+}-j_{f}^{-}}
\end{equation}
while FK's amplitude is a weighed sum over all possible values of
$k_{ef}$, peaking at $k_{ef}=j_{f}=j_{f}^{+}-j_{f}^{-}$(the expression
in brackets is a squared 3j-symbol):
\begin{equation}
W_{e}^{\text{EPRL},\,\gamma<1}=d_{\hat{\iota}_{e}}\prod_{f\in e}\sum_{k_{ef}}d_{k_{ef}}\left[\left(\begin{array}{ccc}
j_{f}^{+} & j_{f}^{-} & k_{ef}\\
j_{f}^{+} & -j_{f}^{-} & j_{f}^{-}-j_{f}^{+}
\end{array}\right)\right]^{2}.
\end{equation}

\end{itemize}

\subsubsection*{Face amplitude $W_{f}$}

Fixing the face amplitude has been an open problem since the inception
of spin foam models, since the structure of Loop Quantum Gravity does
not seem to impose any particular choice for it. It is often associated
with the quantized area of a triangle (see for example \cite{key-72}).
While several choices have been proposed in the literature, the most
common being simply the dimension of the SU(2) representation associated
to the face, $W_{f}=2j_{f}+1$ (indeed, in \cite{key-90} it is argued
it is the correct choice), in the following we shall keep it as general
as possible depending only on the face quantum numbers, $W_{f}\equiv\mu(j_{f})$.\\
\\
For the rest of this study we will use the EPRL prescription, so that
the partition function is (considering a manifold with boundary and
fixed boundary data satisfying Regge-like conditions\cite{key-85}){\small{
\begin{equation}
Z(j_{f_{B}},g_{ve_{B}},n_{ef_{B}})=(-1)^{\chi}\sum_{j_{f}}\prod_{f}\mu(j_{f})\int\prod_{ve}dg_{ve}^{+}dg_{ve}^{-}\int\prod_{ef}dn_{ef}\int\prod_{e}dh_{e}\left(\bigotimes_{f}\mathcal{K}_{f}\right)\circ\left(\bigotimes_{e}\hat{\iota}_{e}(j_{f}^{+}\pm j_{f}^{-},\, n_{ef})\right).\label{eq:part}
\end{equation}
}}It can now be established that the \textit{de facto} variables of
the model are the face SU(2) quantum numbers, Spin(4) elements for
each half-edge $(ve)$ and the coherent state vectors $\left|j_{f}^{+}\pm j_{f}^{-},\, n_{ef}\right\rangle $
for each edge connected to the vertex containing $f$, for each $f$.

\subsection{Path integral formalism}

In order to study the asymptotics of the model, we use the partition
function written in a path integral form,
\begin{equation}
Z=\sum_{c}e^{S[c]}.
\end{equation}
We will review the derivation of this form for the EPRL/FK model\cite{key-93},
but it is worth noting that Bonzom\cite{key-91} has extended the
process for any SFM under some general assumptions.\\
\\
Introducing in (\ref{eq:part}) the expressions for $\hat{\iota}_{e}$
and $\mathcal{K}_{f}$, $\epsilon$-inner products of coherent states
appear. They can be written in terms of the standard Hilbert inner
product by introducing the antilinear structure map $\mathcal{J}:\, V_{k}\rightarrow V_{k}$
defined by
\begin{equation}
\epsilon_{k}(v_{k},v'_{k})=\left\langle \mathcal{J}v_{k}\right|\left.v'_{k}\right\rangle .
\end{equation}
$\mathcal{J}$ has several properties: it commutes with SU(2) group
elements, satisfies $\mathcal{J}^{2}=(-1)^{2k}$ and, since $\mathcal{J}(\vec{n}\cdot\vec{J})=-(\vec{n}\cdot\vec{J})\mathcal{J}$,
it takes a coherent state for the vector $\vec{n}$ to one for $-\vec{n}$.
We should also notice that the orientation requirements described
above (\ref{eq:orient}) are the basis for a supplementary requirement
on the $n_{ef}$, which we will call here the weak gluing condition,
\begin{equation}
\left|n_{ef}\right\rangle _{v}=\mathcal{J}\left|n_{ef}\right\rangle _{v'}\label{eq:weakgluingdef}
\end{equation}
for a tetrahedron that is shared by two vertices. Using this notation
the partition function becomes
\begin{equation}
Z=(-1)^{\chi'}\sum_{j_{f}}\prod_{f}\mu(j_{f})\int\prod_{ve}dg_{ve}^{+}dg_{ve}^{-}\int\prod_{ef}dn_{ef}\prod_{e}dh_{e}\prod_{vf}P_{vf}
\end{equation}
where
\begin{equation}
P_{vf}=\left\langle k_{ef},\mathcal{J}n_{ef}\right|\pi_{k_{ef}}(h_{e}^{-1})C_{j_{f}^{-}j_{f}^{+}}^{k_{ef}}\pi_{j_{f}^{-}}(g_{ev}^{-}g_{ve'}^{-})\pi_{j_{f}^{+}}(g_{ev}^{+}g_{ve'}^{+})C_{k_{e'f}}^{j_{f}^{-}j_{f}^{+}}\pi_{k_{e'f}}(h_{e'})\left|k_{e'f},n_{e'f}\right\rangle 
\end{equation}
can be interpreted as a propagator between two coherent states in
the two edges sharing the face $f$. Now the Clebsch-Gordan maps are
SU(2)-invariant, which means that the $h_{e}$ can be commuted with
the $C$'s into the Spin(4) terms, which take the form $\pi_{j_{f}^{\pm}}(h_{e}^{-1}g_{ev}^{\pm}g_{ve'}^{\pm}h_{e'})$.
The $h_{e}$ can then be eliminated by a change of variables $\tilde{g}_{ve}^{\pm}=g_{ve}^{\pm}h_{e}$,
and the corresponding integrations over them add up to a prefactor
$\text{Vol(SU(2))}^{\#}$.\\
\\
The action of the Clebsch-Gordan maps is simple in the EPRL prescription.
In particular for $\gamma<1$ (the case $\gamma>1$ is slightly more
complicated in analysis but similar in result), we have $k_{ef}=k_{e'f}=j_{f}^{-}+j_{f}^{+}$:
the C-G maps project to the highest spin subspace of $V_{j_{f}^{-}}\otimes V_{j_{f}^{+}}$.
Remembering the property of coherent states that 
\begin{equation}
\left|k,\, n\right\rangle \sim\otimes^{2k}\left|\frac{1}{2},n\right\rangle \equiv\otimes^{2k}\left|n\right\rangle ,
\end{equation}
which is a fully symmetric state and that the highest spin subspace
is precisely the one obtained by full symmetrization, we conclude
that
\begin{equation}
C_{k_{ef}}^{j_{f}^{-}j_{f}^{+}}\left|k_{ef},n_{ef}\right\rangle =\left|k_{ef},n_{ef}\right\rangle =\otimes^{2k}\left|n_{ef}\right\rangle .
\end{equation}
Therefore the propagator simplifies to
\begin{equation}
P_{vf}=\left\langle \mathcal{J}n_{ef}\right|g_{ev}^{-}g_{ve'}^{-}\left|n_{e'f}\right\rangle ^{2j_{f}^{-}}\left\langle \mathcal{J}n_{ef}\right|g_{ev}^{+}g_{ve'}^{+}\left|n_{e'f}\right\rangle ^{2j_{f}^{+}},
\end{equation}
and with some simple algebra we can now write
\begin{equation}
Z=(-1)^{\chi'}\sum_{j_{f}}\mu(j_{f})\int\prod_{ve}dg_{ve}^{+}dg_{ve}^{-}\int\prod_{ef}dn_{ef}e^{S},
\end{equation}
where the ``action'' is
\begin{eqnarray}
S & = & \sum_{f}\sum_{v\in f}2j_{f}^{\pm}\log\left\langle \mathcal{J}n_{ef}\right|g_{ev}^{\pm}g_{ve'}^{\pm}\left|n_{e'f}\right\rangle \nonumber \\
 & \equiv & \sum_{f}S_{f}\label{eq:action}
\end{eqnarray}
Since, by the discussion above, the boundary data are considered to
be fixed for the ``path-integral'' approach, while only the interior
data are dynamical, it is important to separate the action into its
boundary and interior parts, $S\text{=}S_{I}+S_{B}=\sum_{f_{I}}S_{f}+\sum_{f_{B}}S_{f}$.
In section 3 we will see how the action here written can be related
to that of Regge calculus in the large-j regime, the base point of
the asymptotics discussion.

\section{Asymptotics: general considerations and past work}

The semiclassical limit in quantum gravity is commonly taken in the
literature as the limit of large areas, since the discrete area spectrum
of LQG is asymptotically indistiguishable from the continuous classical
spectrum when the corresponding quantum number $j_{f}$ is large (i.e.
$\frac{\Delta j}{j}\underset{j\rightarrow\infty}{\rightarrow}0$)
. Mathematically this is imposed by making the transformation $j_{f}\rightarrow\lambda j_{f},\,\forall f$
in the regime $\lambda\rightarrow\infty$. For the EPRL model this
means that its action is proportional to $\lambda$, so that the partition
function is (roughly) of the form
\begin{equation}
I_{\lambda}=\int d^{n}z\, g(z)e^{\lambda F(z)},\,\lambda\rightarrow\infty.\label{eq:statphaseone}
\end{equation}
This suggests the use of the stationary phase method to derive an
approximation of $I_{\lambda}$ in the large $\lambda$ limit.

\subsection{The stationary phase method}

The main principle of the stationary phase method is that due to the
large argument of the exponential in the integrand, the contributions
to the integral near certain \emph{critical points} are much larger
than everywhere else, and the integral can be estimated by considering
the function only near those points. Critical points are given by
the following conditions:
\begin{itemize}
\item $\Re(F(z))$ is at its absolute maximum, so that $\left|e^{\lambda F(z)}\right|$
is maximized;
\item the oscillation is minimized, i.e. the variation of $\arg\left(e^{\lambda F(z)}\right)$
in a neighbourhood of the point in question is the slowest. At a first
order level this is obtained by extremizing the action, i.e. $\partial_{i}f(z)=0,\,\forall i.$,
so that the variation of $\Im(F(z))$ near a critical point $z_{0}$
is at least second order in $z-z_{0}$, rather than first. 
\end{itemize}
While not a rigorous proof (see \cite{key-96,key-97} for more detailed
mathematical treatment), the essentials of the method can be understood
with the following argument. That we need to maximize the real part
of $F(z)$ should be obvious in the large $\lambda$ regime, so assume
in the following that $F(z)=if(z),\, f\in\mathbb{R}$, and for simplicity
$g(z)\equiv1$ (the only condition on $g$ is that it allows for convergence
of the integral, which won't be a problem in the cases we are interested
in considering). Take a Taylor expansion of $f$ around an arbitrary
point $z_{0}$:
\begin{eqnarray}
f(z) & \approx & f(z_{0})+\left.\frac{\partial f}{\partial z^{i}}\right|_{z_{0}}(z-z_{0})^{i}+\frac{1}{2}\left.\frac{\partial^{2}f}{\partial z^{i}\partial z^{j}}\right|_{z_{0}}(z-z_{0})^{i}(z-z_{0})^{j}\nonumber \\
 & + & \frac{1}{3!}\left.\frac{\partial^{3}f}{\partial z^{i}\partial z^{j}\partial z^{k}}\right|_{z_{0}}(z-z_{0})^{i}(z-z_{0})^{j}(z-z_{0})^{k}+\mathcal{O}(z^{4})\nonumber \\
 & \equiv & f(z_{0})+D_{i}(z_{0})(z-z_{0})^{i}+H_{ij}(z_{0})(z-z_{0})^{i}(z-z_{0})^{j}\nonumber \\
 & + & T_{ijk}(z_{0})(z-z_{0})^{i}(z-z_{0})^{j}(z-z_{0})^{k}+\mathcal{O}(z^{4})
\end{eqnarray}
The stationary phase method assumes that when $z_{0}$ are critical
points, the integral (\ref{eq:statphaseone}) is estimated by the
formula
\begin{equation}
I_{\lambda}\approx\int dz_{0}\int_{U(z_{0})}d^{n}z\, e^{i\lambda f(z)}
\end{equation}
where $U(z_{0})$ is a neighbourhood of $z_{0}$. Now suppose we only
took the first order term in the Taylor expansion of $f$. Then 
\begin{eqnarray}
I_{\lambda}^{1} & \approx & \int dz_{0}\int_{U(z_{0})}d^{n}z\,\exp[i\lambda(f(z_{0})+D_{i}(z_{0})(z-z_{0})^{i})]\nonumber \\
 & = & \int dz_{0}\exp[i\lambda(f(z_{0})+D_{i}(z_{0})z_{0}^{i})]\int_{U(z_{0})}d^{n}z\, e^{i\lambda D_{i}(z_{0})z^{i}}
\end{eqnarray}
If we further assume that the contribution away from a critical point
is (after taking the Taylor approximation) so small that the integral
above can be extended to the whole $z$-space, the integral over $z$
is directly related to the delta ``function'':
\begin{equation}
\int d^{n}z\, e^{i\lambda D_{i}(z_{0})z^{i}}=\frac{1}{2\pi\lambda}\delta(D_{i}(z_{0}))
\end{equation}
in this extremely crude approximation, divergences show up when $D_{i}(z_{0})=0$.
While this points to the necessity of refining the method, which happens
by taking the Taylor expansion to second order (enough in most applications),
it also serves as a very simple justification that the contributions
of points $z_{0}$ satisfying $D_{i}(z_{0})=0$ are dominant, justifying
the definition of critical point above. Taking the second order expansion
of $f$, then, we get the more accurate formula
\begin{eqnarray}
I_{\lambda}^{2} & = & \int d^{n}z_{0}\exp[i\lambda(f(z_{0})+D_{i}(z_{0})z_{0}^{i})]\int d^{n}z\,\exp[i\lambda(D_{i}(z_{0})z^{i})+H_{ij}(z_{0})(z-z_{0})^{i}(z-z_{0})^{j}]\prod_{i}\delta(D_{i}(z_{0}))\nonumber \\
 & = & \int_{\Sigma_{C}}d^{n}z_{0}e^{i\lambda f(z_{0})}\int d^{n}z\, e^{i\lambda H_{ij}(z_{0})(z-z_{0})^{i}(z-z_{0})^{j}}\label{eq:secondorder}
\end{eqnarray}
where $\Sigma_{C}$, the \emph{critical surface},\emph{ }is the hypersurface%
\footnote{The critical surface is in fact a submanifold of $z$-space iff $\det H_{r}(z_{0})\neq0\,\forall z_{0}\in\Sigma_{C}$.%
} of $z$-space formed by all critical points. Using analytic continuation
of the standard formula $\int d^{n}x\, e^{-\frac{1}{2}A_{\alpha\beta}x^{\alpha}x^{\beta}}=\sqrt{\frac{(2\pi)^{n}}{det\, A}}$
to complex $A$, we can solve the integral over $z$:
\begin{equation}
\int d^{n}z\, e^{i\lambda H_{ij}(z_{0})(z-z_{0})^{i}(z-z_{0})^{j}}=\left(\frac{2\pi}{i\lambda}\right)^{n/2}\frac{1}{\sqrt{\det H_{r}(z_{0})}}
\end{equation}
where $H_{r}$ is the restriction of $H$ to the orthogonal complement
of its null space, as the conditions imposed on the $z_{0}$ constrain
some degrees of freedom of $H$.

\subsection{EPRL asymptotics: the reconstruction theorem}

In the context of state sum models the critical point equations can
be interpreted as classical equations of motion for the interior variables
of the simplicial complex (boundary data is fixed). Considering the
action (\ref{eq:action}) for the Euclidean EPRL model with $0<\gamma<1$,
the equations of motion are
\begin{eqnarray}
\Re(S_{I}) & = & R_{max}\label{eq:Re-1}\\
\delta_{g_{ve}}S_{I} & = & 0\label{eq:g-1}\\
\delta_{n_{ef}}S_{I} & = & 0\label{eq:n-1}\\
\delta_{j_{f_{I}}}S_{I} & = & 0\label{eq:x-1-1}
\end{eqnarray}
Or are they? (\ref{eq:x-1-1}) in particular has rarely been considered
in existing literature. The main reason is simple - unlike the other
spin foam variables in play, the $j_{f}\in\frac{\mathbb{N}}{2}$ are
discrete, and it is unclear whether there is an extension of the stationary
phase method applying to sums over general discrete variables. The
only work in this direction that we are aware of is Lachaud's\cite{key-98}
results for sums over finite fields, which is in general not the case
of the $j_{f}$ sums.\\
\\
The other equations of motion can be written explicitly, and are as
follows:
\begin{itemize}
\item (\ref{eq:Re-1}) gives the\emph{ gluing condition:} $R(g_{ve}^{\pm})\vec{n}_{ef}=-R(g_{ve'}^{\pm})\vec{n}_{e'f}$,
where $R(g)$ is the rotation matrix associated to $g$ by the 2-1
surjective homomorphism $\text{SU(2)}\rightarrow\text{SO(3)}$;
\item (\ref{eq:g-1}) gives the\emph{ closure} \emph{condition: $\sum_{f\in e}\sum_{\pm}2j_{f}^{\pm}\epsilon_{ef}(v)R(g_{ve}^{\pm})\vec{n}_{ef}=0$},
where $\epsilon_{ef}(v)$ is defined to be 1 if the orientation of
$f$ agrees with the one induced from $e$ according to (\ref{eq:orient}),
and -1 otherwise. $\epsilon_{ef}(v)$ are also subject to the orientation
conditions, $\epsilon_{ef}(v')=-\epsilon_{ef}(v)=-\epsilon_{e'f}(v')$.
\item if the previous two conditions are met, (\ref{eq:n-1}) is automatically
satisfied.
\end{itemize}
The main existing result for EPRL asymptotics is the \emph{reconstruction
theorem}, proven originally by Barrett et al\cite{key-85} for the
case of one single 4-simplex, and more recently extended by Han and
Zhang\cite{key-82,key-83}%
\footnote{Han and Zhang developed their results for both the Euclidean and Lorentzian
signature versions of the EPRL model. We will focus on Euclidean signature
for this paper.%
} for a general simplicial complex with boundary. Essentially, the
reconstruction theorem states that given a set of boundary data satisfying
a number of conditions guaranteeing their geometricity, called ``Regge-like'',
and non-degenerate interior spin foam variables $j_{f},\, g_{ve},\, n_{ef}$
satisfying the equations of motion, then it is possible to construct
a classical, non-degenerate geometry which matches them and is unique
up to global symmetries. The proof is constructive and involves defining
bivectors $X_{ef}(j_{f},n_{ef})$ which are interpreted as area bivectors
of the discrete geometry, while the $g_{ve}$ are identified with
the spin connection (in both cases up to sign factors). Additionally,
the Regge deficit angles $\Theta_{f}$ can be identified within the
bivector formalism, such that the semiclassical action is found to
be
\begin{equation}
S=\sum_{f}i\epsilon\left[j_{f}\mathcal{N}_{f}\pi-\gamma j_{f}\text{sign}(V_{4})\Theta_{f}\right]\label{eq:HZresult}
\end{equation}
where $\mathcal{N}_{f}\in\mathbb{N}$ and $V_{4}$ is the 4-volume
of the connected component of the discrete manifold that contains
$f$ , its sign depending on the orientation induced from spin foam
variables. Since the first term is a half-integer times $i\pi$ and
only gives a $\pm$ sign when exponentiated, it is mostly ignored,
so this ``classical'' form for $S$ bears an uncanny resemblance
to the discrete Einstein-Hilbert action in Regge calculus\cite{key-92}:
\begin{equation}
S_{Regge}=\sum_{f}A_{f}\Theta_{f}
\end{equation}
where $A_{f}$ is the area of the triangle $f$, which coincides with
$\gamma j_{f}$ in the reconstructed geometry.

\subsection{The j-equation and the Flatness Problem}

Given (\ref{eq:HZresult}), it is readily seen how the j-equation
(\ref{eq:x-1-1}) was the original motivation to the ``flatness problem''
mentioned by Freidel and Conrady\cite{key-94} and later Bonzom\cite{key-91}.
The result shows that the EPRL action (\ref{eq:action}) can be written
as
\begin{eqnarray*}
S & = & \sum_{f_{I}}j_{f}\tilde{\Theta}_{f}(g_{ve},n_{ef})
\end{eqnarray*}
where $\tilde{\Theta}_{f}$ is a quantity that is proportional, in
the semiclassical limit, to the Regge-like deficit angle, $\tilde{\Theta}_{f}\underset{\lambda\rightarrow\infty}{\rightarrow}\pm\gamma\Theta_{f}$.
If we were to ignore the discreteness of the $j_{f}$ and carry out
the derivation as if it were continuous, the j-equation would be simply
$\tilde{\Theta}_{f}=0,\forall f$, therefore showing that the classical
geometries reproduced by the model are restricted to be flat - a result
that puts the model in question, since GR in four dimensions admits
curved spacetime solutions. However, the applicability of this equation
is questionable, not only because of the issues with the discreteness
of $j_{f}$, but due to an ambiguity in the way the semiclassical
limit is taken - taking the limit of large $j_{f}$, while at the
same time summing over them. In the following we consider a slight
reformulation.\\
\\
Assume that in the semiclassical limit the boundary face quantum numbers
are given by $j_{f_{B}}=\lambda j\text{'}_{f_{B}},\,\forall f_{B}$
where $j'_{f_{B}}\in\frac{\mathbb{N}}{2}$ and $\lambda\rightarrow\infty$.
Then, define new interior variables $x_{f_{I}}=\frac{j_{f_{I}}}{\lambda}\in\frac{\mathbb{N}}{2\lambda}$
(and $x_{f_{I}}^{\pm}$ accordingly). The partition function then
takes the form
\begin{equation}
Z(\lambda j\text{'}_{f_{B}},\, g_{ve_{B}},\, n_{ef_{B}})=\sum_{x_{f_{I}}}\int\prod_{ve}dg_{ve}\int\prod_{ef}dn_{ef}e^{i\lambda(S_{I}+S_{B})}
\end{equation}
with
\begin{eqnarray}
S_{I} & =-i & \sum_{f_{I}}\sum_{v\in f}\sum_{\pm}2x_{f}^{\pm}\log\left\langle \mathcal{J}n_{ef}\right|g_{ev}^{\pm}g_{ve'}^{\pm}\left|n_{e'f}\right\rangle \equiv\sum_{f_{I}}x_{f_{I}}\tilde{\Theta}_{f_{I}}(g_{ve},n_{ef})\nonumber \\
S_{B} & =-i & \sum_{f_{B}}\sum_{v\in f}\sum_{\pm}2j_{f}^{'\pm}\log\left\langle \mathcal{J}n_{ef}\right|g_{ev}^{\pm}g_{ve'}^{\pm}\left|n_{e'f}\right\rangle 
\end{eqnarray}
(we factor out $i$ to explicit the fact that the argument of the
exponential becomes pure imaginary when the gluing condition is satisfied).
With this prescription, we don't have to assume anything about the
$x_{f_{I}}$'s, eliminating ambiguities, and the dependence of the
partition function on $\lambda$ is completely explicit. Additionally,
we can propose a workaround to the discreteness issue, consisting
of a continuum approximation for the $x_{f}$. Since the $\Delta x_{f_{I}}=\frac{1}{2\lambda}$
tend to zero for large $\lambda$, it makes sense to consider replacing
the sum over $x_{f}$ by an integral:
\begin{equation}
\frac{1}{\Delta x_{f_{I}}}\sum_{x_{f_{I}}}f(x_{f_{I}})\Delta x_{f_{I}}\approx\frac{1}{\Delta x_{f_{I}}}\int_{0}^{\infty}f(x_{f_{I}})dx_{f_{I}}
\end{equation}
and therefore the ``semiclassical'' partition function would be
\begin{equation}
Z_{SC}(\lambda j\text{'}_{f_{B}})=\left(2\lambda\right)^{\#f_{I}}\int\prod_{f_{I}}dx_{f}\int\prod_{(ve)_{I}}dg_{ve}\int\prod_{(ef)_{I}}dn_{ef}e^{i\lambda(S_{I}+S_{B})}
\end{equation}
Of course, one must be careful with the errors incurring from this
approximation, which is essentially the rectangle method of numerical
integration ``done backwards''. It can be shown%
\footnote{Consider the difference $\int_{x_{0}}^{x_{0}+\Delta x}f(x)dx-f(x_{0})\Delta x.$
For $\Delta x=1/2\lambda$ the difference is of order $1/\lambda^{2}$.
In practical semiclassical calculations the integral will not extend
to infinity because triangle inequalities limit the maximum value
of $j$. The cutoff will be of order $\lambda$, so the error in approximating
the sum by an integral is of order 1/$\lambda$.%
} that the difference between the sum and the integral is of order
$\frac{1}{\lambda}$, making the continuum approximation unreliable
to compute any quantum corrections to the zero-order, $\lambda=\infty$
results. It could still be argued that that it can be used safely
in the zero-order situation, but we will try to progress as much as
possible without using it. The problem is to estimate the integral
\begin{equation}
\sum_{j_{f}}\mu(j_{f})\int dY\, e^{\sum_{f}i\lambda x_{f}\tilde{\Theta}_{f}(Y)}
\end{equation}
where we used $Y$ as short for the set of $g_{ve},\, n_{ef}$ integration
variables. Using the stationary phase method for the integral over
$Y$, we obtain{\small{
\begin{equation}
\int dY\, e^{\sum_{f}i\lambda x_{f}\tilde{\Theta}_{f}(Y)}\approx\int_{\Sigma_{C}(x_{f})}dY_{C}\prod_{f}e^{i\lambda x_{f}\tilde{\Theta}_{f}(Y_{C})}\left(\frac{-2\pi i}{\lambda}\right)^{\#Y_{C}/2}\frac{1}{\sqrt{\det\left[\sum_{f}x_{f}H_{r}^{f}(Y_{C})\right]}}
\end{equation}
}}where $Y_{C}$ are the critical points that solve the equations
of motion, and $\Sigma_{C}$ the submanifold of $Y$-space they form.
Ideally, if we use the continuum approximation, we could think of
reversing order of integration and doing the $x$ integral first,
but this is not possible for the general case because not only there
is an $x$ dependence on the determinant factor, which is a priori
arbitrary, but due to the closure condition the critical surface $\Sigma_{C}$
also depends on $x$. This makes the integral seemingly intractable
without further assumptions. There are some heuristic considerations
that can be made on this form of $Z$ that lead to something suggestive
of the flatness problem, but the apparent ``dead end'' we reach
here leads us to consider a concrete example in which a full calculation
is possible, the $\Delta_{3}$ manifold studied in section 4.\\
\\
More recently, a different approach to asymptotics devised by Hellmann
and Kaminski\cite{key-101} derived a result similar to the flatness
problem. Their main idea is to introduce the concept of wavefront
sets for a distribution, which are designed with asymptotics in mind
and represent the subspace of phase space where the distribution is
peaked in the limit of large $\lambda$. The wavefront sets of partition
functions of various models like BC and EPRL can be written using
the holonomy (or operator) representation of spin foams\cite{key-102}
and their main result regarding asymptotics is an accidental curvature
constraint acting on the deficit angles $\Theta_{f}$,
\begin{equation}
\gamma\Theta_{f}=0\,\mod\,2\pi,
\end{equation}
which is not strictly flatness (the dependence on the Immirzi parameter
is somewhat puzzling) but still a worrying result in terms of the
accuracy of the theory's asymptotics in respect to Einstein theory.
It is noteworthy that for the BC model, which can essentially be obtained
form EPRL by taking the limit $\gamma\rightarrow\infty$, the wavefront
approach leads to an exact flatness constraint.

\section{An example: $\Delta_{3}$}

In the following we will attempt to compute the asymptotic EPRL partition
function for the case of the three 4-simplex manifold $\Delta_{3}$,
which is represented in the figure below together with its 2-complex
dual. This particular manifold is chosen as a simple example of a
semiclassical calculation, since it has only one interior face $f_{I}$.
Therefore, assuming the boundary data are fixed, Regge-like, and non-degenerate,
the classical Euclidean geometry of $\Delta_{3}$ is completely determined
by the area $j=\lambda x$ and the deficit angle $\Theta$ of $f_{I}$,
two quantities that are easily seen to be completely determined by
the boundary geometry. We will now define the EPRL model in this triangulation.

\begin{center}
\includegraphics[scale=0.7]{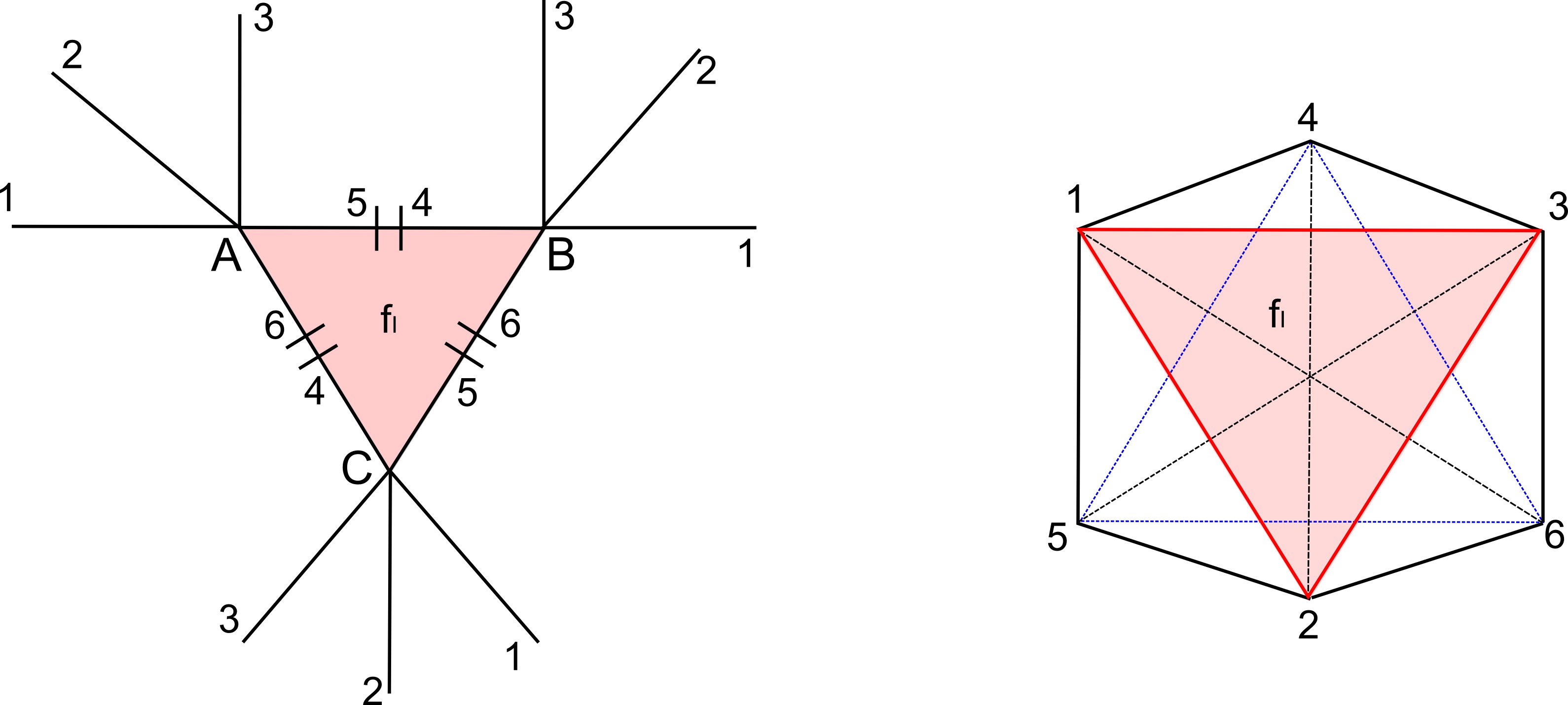}
\par\end{center}

\noindent Boundary faces are notated $f_{ij}^{v},\, i,j\in\{1,...,5\}$
where $f_{ij}^{v}$ is the triangle that does not contain the points
$i,j$ of the 4-simplex $v$ it belongs to, and has the area variable
$x_{ij}^{v}$. Edges are labelled $e_{k}^{v},\, k\in\{1,...,5\}$
and $e_{k}^{v}$ is the tetrahedron that does not contain the point
$k$ of $v$. We will call the $n_{ef}$ as $\left|n_{e,f}\right\rangle _{v},\, v\in\{A,B,C\}$
for clarity, while the interior $g_{ve}$ are labelled $g_{A5}$,
$g_{A6}$, $g_{B5}$, $g_{B6}$, $g_{C5}$, $g_{C6}$ according to
the figure. The partition function is (proportional to, with extra
pre-factors not being of importance in the analysis)
\begin{equation}
Z=\sum_{x=j/\lambda}\frac{\mu(\lambda x)}{x^{\#Y_{C}}}\int_{\Sigma_{C}(x)}dY_{c}\frac{e^{i\lambda x\tilde{\Theta}(Y_{C})}}{\sqrt{\det H_{r}(Y_{C})}}
\end{equation}
noting that the dimension $\#Y$ of $Y$-space is that of 12 copies
of $S^{3}$ associated to the interior $g_{ve}$ and other 6 copies
associated to the interior $n_{ef}$. The dimension $\#Y_{C}$ of
the critical surface is the number of degrees of freedom unconstrained
by the equations of motion.

\subsection{Solving the equations of motion}

\noindent We will now study the equations of motion for $\Delta_{3}$.
For starters, $n_{ef}^{v}$ and $n_{ef}^{v'}$ are related by the
weak gluing equations (\ref{eq:weakgluingdef}): 
\begin{eqnarray}
\left|n_{6,56}\right\rangle _{A} & = & \mathcal{J}\left|n_{4,45}\right\rangle _{C}\nonumber \\
\left|n_{5,45}\right\rangle _{C} & = & \mathcal{J}\left|n_{6,46}\right\rangle _{B}\nonumber \\
\left|n_{4,46}\right\rangle _{B} & = & \mathcal{J}\left|n_{5,56}\right\rangle _{A}\label{eq:weakgluing}
\end{eqnarray}
We can choose a simpler notation for the interior $n_{ef}$ so that
(\ref{eq:weakgluing}) reads
\begin{eqnarray}
\left|n_{AC}\right\rangle  & = & \mathcal{J}\left|n_{CA}\right\rangle \nonumber \\
\left|n_{CB}\right\rangle  & = & \mathcal{J}\left|n_{BC}\right\rangle \nonumber \\
\left|n_{BA}\right\rangle  & = & \mathcal{J}\left|n_{AB}\right\rangle 
\end{eqnarray}
Stationary phase computation on the $g,\, n$ integrals results in
6 interior gluing conditions,
\begin{eqnarray}
R(g_{C4}^{\pm})\triangleright\vec{n}_{CA} & = & -R(g_{C5}^{\pm})\triangleright\vec{n}_{CB}\nonumber \\
R(g_{B6}^{\pm})\triangleright\vec{n}_{BC} & = & -R(g_{B4}^{\pm})\triangleright\vec{n}_{BA}\nonumber \\
R(g_{A5}^{\pm})\triangleright\vec{n}_{AB} & = & -R(g_{A6}^{\pm})\triangleright\vec{n}_{AC}\label{eq:gluing}
\end{eqnarray}
36 interior-boundary gluing conditions,
\begin{eqnarray}
R(g_{A5}^{\pm})\triangleright\vec{n}_{5,i5}^{A} & = & -R(g_{Ai}^{\pm})\triangleright\vec{n}_{i,i5}^{A}\nonumber \\
R(g_{A6}^{\pm})\triangleright\vec{n}_{6,i6}^{A} & = & -R(g_{Ai}^{\pm})\triangleright\vec{n}_{i,i6}^{A}\nonumber \\
R(g_{B6}^{\pm})\triangleright\vec{n}_{6,i6}^{B} & = & -R(g_{Bi}^{\pm})\triangleright\vec{n}_{i,i6}^{B}\nonumber \\
R(g_{B4}^{\pm})\triangleright\vec{n}_{4,i4}^{B} & = & -R(g_{Bi}^{\pm})\triangleright\vec{n}_{i,i4}^{B}\nonumber \\
R(g_{C4}^{\pm})\triangleright\vec{n}_{4,i4}^{C} & = & -R(g_{Ci}^{\pm})\triangleright\vec{n}_{i,i4}^{C}\nonumber \\
R(g_{C5}^{\pm})\triangleright\vec{n}_{5,i5}^{C} & = & -R(g_{Ci}^{\pm})\triangleright\vec{n}_{i,i5}^{C},\, i\in\{1,2,3\}\label{eq:boundarygluing}
\end{eqnarray}
and 6 closure conditions,
\begin{eqnarray}
x\left[(1+\gamma)R(g_{C4}^{+})+(1-\gamma)R(g_{C4}^{-})\right]\triangleright\vec{n}_{CA}+b.t._{(C+)} & = & 0\nonumber \\
x\left[(1+\gamma)R(g_{A6}^{+})+(1-\gamma)R(g_{A6}^{-})\right]\triangleright\vec{n}_{AC}+b.t._{(A+)} & = & 0\nonumber \\
x\left[(1+\gamma)R(g_{B6}^{+})+(1-\gamma)R(g_{B6}^{-})\right]\triangleright\vec{n}_{BC}+b.t._{(B+)} & = & 0\label{eq:closure+}\\
-x\left[(1+\gamma)R(g_{C5}^{+})+(1-\gamma)R(g_{C5}^{-})\right]\triangleright\vec{n}_{CB}+b.t._{(C-)} & = & 0\nonumber \\
-x\left[(1+\gamma)R(g_{A5}^{+})+(1-\gamma)R(g_{A5}^{-})\right]\triangleright\vec{n}_{AB}+b.t._{(A-)} & = & 0\nonumber \\
-x\left[(1+\gamma)R(g_{B4}^{+})+(1-\gamma)R(g_{B4}^{-})\right]\triangleright\vec{n}_{BA}+b.t._{(B-)} & = & 0\label{eq:closure-}
\end{eqnarray}
where the $b.t.$ represents terms depending exclusively on boundary
variables. Indeed, the closure conditions contain sums over edges
in each vertex, so each of them contains exactly one term corresponding
to the interior edge, and the rest of the sum depends on the boundary
edge variables. The boundary terms are labelled by the edges they
pertain to.\\
\\
First off, we will note that Eqs. (\ref{eq:boundarygluing}) determine
all the interior $g_{ve}$ uniquely in terms of boundary data. Indeed,
consider the first equation referring to $g_{A5}^{\pm}$. The only
term in this equation that is not a boundary variable is $R(g_{A5}^{\pm})$,
and the indices 1,2,3 can be grouped in a matrix form equation:
\begin{equation}
R(g_{A5}^{\pm})\triangleright\underbrace{\left[\begin{array}{ccc}
\vec{n}_{5,15}^{A} & \vec{n}_{5,25}^{A} & \vec{n}_{5,35}^{A}\end{array}\right]}_{\equiv N_{A5}}=-\underbrace{\left[\begin{array}{ccc}
R(g_{A1}^{\pm})\triangleright\vec{n}_{1,15}^{A} & R(g_{A2}^{\pm})\triangleright\vec{n}_{2,25}^{A} & R(g_{A2}^{\pm})\triangleright\vec{n}_{2,25}^{A}\end{array}\right]}_{\equiv V_{A5}^{\pm}}
\end{equation}
Note that the non-degeneracy assumption on the boundary data implies
that, since all tetrahedra are non-degenerate, any set of three out
of the four $\vec{n}_{ef}$ that define a tetrahedron must be linearly
independent. This means that $N_{A5}$ is invertible in the equation
above, which can then immediately be solved:
\begin{equation}
R(g_{A5}^{\pm})=-N_{A5}^{-1}V_{A5}^{\pm}
\end{equation}
and similar solutions are derived for the remaining $g_{ve}.$ This
result means that the purely interior gluing conditions (\ref{eq:gluing}),
if consistent (consistency should be guaranteed by the boundary data
being Regge-like), are redundant, however we will analyse them together
with the closure conditions in the following, as they have valuable
physical content for the problem.\\
\\
It is possible to eliminate three of the closure equations by using
the gluing ones: indeed, substituting (\ref{eq:gluing}) on (\ref{eq:closure-}),
we obtain (\ref{eq:closure+}) while being forced to impose that $b.t._{(A+)}=-b.t._{(A-)}$
(and similar for the $B\pm$ and $C\pm$ boundary terms). Conditions
on boundary variables are not problematic if they can be related to
the equations for Regge-like data. To elaborate on this and to properly
solve the closure conditions we need to specify the boundary data.
The equations (\ref{eq:closure+}) in their full form are
\begin{eqnarray}
\left[(1+\gamma)R(g_{C4}^{+})+(1-\gamma)R(g_{C4}^{-})\right]\triangleright\left(x\vec{n}_{CA}+x_{41}^{C}\vec{n}_{4,41}^{C}+x_{42}^{C}\vec{n}_{4,42}^{C}+x_{43}^{C}\vec{n}_{4,43}^{C}\right) & = & 0\nonumber \\
\left[(1+\gamma)R(g_{B6}^{+})+(1-\gamma)R(g_{B6}^{-})\right]\triangleright\left(x\vec{n}_{BC}+x_{61}^{B}\vec{n}_{6,61}^{B}+x_{62}^{B}\vec{n}_{6,62}^{B}+x_{63}^{B}\vec{n}_{6,63}^{B}\right) & = & 0\nonumber \\
\left[(1+\gamma)R(g_{A5}^{+})+(1-\gamma)R(g_{A5}^{-})\right]\triangleright\left(x\vec{n}_{AB}+x_{51}^{A}\vec{n}_{5,51}^{A}+x_{52}^{A}\vec{n}_{5,52}^{A}+x_{53}^{A}\vec{n}_{5,53}^{A}\right) & = & 0
\end{eqnarray}
The solution of these equations is simple to obtain, noting that they
are of the form $M\triangleright\vec{v}=0$, a condition satisfied
if and only if $\vec{v}=0$ or $M$ has a vanishing determinant. The
second possibility can be ruled out, though, by proving that $M=(1+\gamma)G+(1-\gamma)H$
has nonzero determinant for all $G,H\in SO(3)$ and $0<\gamma<1.$
Proof starts with noting that $(\det M)^{2}=\det(M^{t}M)$. It is
possible to get a general expression for $\det(M^{t}M)$: 
\begin{eqnarray}
M^{t}M & = & \left[(1+\gamma)G^{t}+(1-\gamma)H^{t}\right]\left[(1+\gamma)G+(1-\gamma)H\right]\nonumber \\
 & = & 2(1+\gamma^{2})\boldsymbol{1}+(1-\gamma^{2})(G^{t}H+H^{t}G)\nonumber \\
 & = & 2(1+\gamma^{2})\boldsymbol{1}+(1-\gamma^{2})(A+A^{t})
\end{eqnarray}
defining $A\equiv G^{t}H\in SO(3)$. We can compute the determinant
in a basis where $A+A^{t}$ is diagonal - note that the identity matrix
is basis-invariant and $A+A^{t}$ is a symmetric real matrix, hence
diagonalizable. To do so we need its eigenvalues, which can be found
using one of the several possible parameterizations of $SO(3)$. Here
we use a parameterization by Janaki and Rangarajan\cite{key-99}:
\begin{equation}
A=\left[\begin{array}{ccc}
\cos\theta_{1}\cos\theta_{2} & \sin\theta_{1}\cos\theta_{3}-\cos\theta_{1}\sin\theta_{2}\sin\theta_{3} & \sin\theta_{1}\sin\theta_{3}+\cos\theta_{1}\sin\theta_{2}\cos\theta_{3}\\
-\sin\theta_{1}\cos\theta_{2} & \cos\theta_{1}\cos\theta_{3}+\sin\theta_{1}\sin\theta_{2}\sin\theta_{3} & \cos\theta_{1}\sin\theta_{3}-\sin\theta_{1}\sin\theta_{2}\cos\theta_{3}\\
-\sin\theta_{2} & -\cos\theta_{2}\sin\theta_{3} & \cos\theta_{2}\cos\theta_{3}
\end{array}\right]
\end{equation}
where $\theta_{i}\in[0,2\pi]$ are angles for simple rotations. $A+A^{t}$
can then be diagonalized, being a symmetric real matrix. There is
a basis in which $A+A^{t}=\left[\begin{array}{ccc}
a\\
 & b\\
 &  & c
\end{array}\right],$ where
\begin{eqnarray}
a & = & 2\nonumber \\
b=c & = & \sin\theta_{1}\sin\theta_{2}\sin\theta_{3}+\cos\theta_{1}(\cos\theta_{2}+\cos\theta_{3})+\cos\theta_{2}\cos\theta_{3}-1
\end{eqnarray}
are its eigenvalues. In this basis,
\begin{eqnarray}
M^{t}M & = & 2(1+\gamma^{2})\left[\begin{array}{ccc}
1\\
 & 1\\
 &  & 1
\end{array}\right]+(1-\gamma^{2})\left[\begin{array}{ccc}
2\\
 & b\\
 &  & b
\end{array}\right]\nonumber \\
 & = & \left[\begin{array}{ccc}
4\\
 & 2(1+\gamma^{2})+b(1-\gamma^{2})\\
 &  & 2(1+\gamma^{2})+b(1-\gamma^{2})
\end{array}\right]
\end{eqnarray}
so that $(\det M)^{2}=4\left[2(1+\gamma^{2})+b(1-\gamma^{2})\right]^{2}$.
Therefore,
\begin{equation}
\det M=0\Leftrightarrow b=-2\frac{1+\gamma^{2}}{1-\gamma^{2}}
\end{equation}
It is straightforward to verify that $-2\leq b\leq2$ for all values
of $\theta_{i}$, which makes the above condition impossible in the
$0<\gamma<1$ range we are working on. Hence, $M$ is always invertible
in the conditions of our study, and the closure conditions are simplified:
\begin{eqnarray}
x\vec{n}_{CA}+x_{41}^{C}\vec{n}_{4,41}^{C}+x_{42}^{C}\vec{n}_{4,42}^{C}+x_{43}^{C}\vec{n}_{4,43}^{C} & = & 0\nonumber \\
x\vec{n}_{BC}+x_{61}^{B}\vec{n}_{6,61}^{B}+x_{62}^{B}\vec{n}_{6,62}^{B}+x_{63}^{B}\vec{n}_{6,63}^{B} & = & 0\nonumber \\
x\vec{n}_{AB}+x_{51}^{A}\vec{n}_{5,51}^{A}+x_{52}^{A}\vec{n}_{5,52}^{A}+x_{53}^{A}\vec{n}_{5,53}^{A} & = & 0
\end{eqnarray}
Notice that these are precisely the necessary and sufficient conditions
for the 3 tetrahedra of $\Delta_{3}$ that contain the interior face
$f$ to be geometrical in the Euclidean sense, which shows that the
large areas limit for this manifold imposes a discrete classical geometry
on it. Also, the partition function is considerably simplified, since
$x$ and all the interior $\vec{n}_{ef}$ are fixed:
\begin{eqnarray}
x & = & \left|x_{41}^{C}\vec{n}_{4,41}^{C}+x_{42}^{C}\vec{n}_{4,42}^{C}+x_{43}^{C}\vec{n}_{4,43}^{C}\right|\nonumber \\
\vec{n}_{BA} & = & -\frac{x_{41}^{C}\vec{n}_{4,41}^{C}+x_{42}^{C}\vec{n}_{4,42}^{C}+x_{43}^{C}\vec{n}_{4,43}^{C}}{\left|x_{41}^{C}\vec{n}_{4,41}^{C}+x_{42}^{C}\vec{n}_{4,42}^{C}+x_{43}^{C}\vec{n}_{4,43}^{C}\right|}
\end{eqnarray}
and similarly for $\vec{n}_{AC}$ and $\vec{n}_{CB}$. In particular,
note that the j-equation (\ref{eq:x-1-1}) seems not to apply in this
example: $x$ is fixed in terms of boundary data by the gluing/closure
conditions, without need of an extra equation for it. Note that the
other two closure conditions also give expressions for $x$, leading
to additional constraints on boundary data:
\begin{equation}
\left|x_{13}^{A}\vec{n}_{1,13}^{A}+x_{14}^{A}\vec{n}_{1,14}^{A}+x_{15}^{A}\vec{n}_{1,15}^{A}\right|=\left|x_{13}^{B}\vec{n}_{1,13}^{B}+x_{14}^{B}\vec{n}_{1,14}^{B}+x_{15}^{B}\vec{n}_{1,15}^{B}\right|=\left|x_{13}^{C}\vec{n}_{1,13}^{C}+x_{14}^{C}\vec{n}_{1,14}^{C}+x_{15}^{C}\vec{n}_{1,15}^{C}\right|.
\end{equation}
Additionally, the relations between (\ref{eq:closure+}) and (\ref{eq:closure-})
make it so that
\begin{eqnarray}
\vec{n}_{CA} & = & -\vec{n}_{CB}\nonumber \\
\vec{n}_{BC} & = & -\vec{n}_{BA}\nonumber \\
\vec{n}_{AC} & = & -\vec{n}_{AB}
\end{eqnarray}
and together with weak gluing, we obtain that $\vec{n}_{AB}=\vec{n}_{BC}=\vec{n}_{CA}\equiv\vec{n}$.
The partition function is now reduced to
\begin{equation}
Z=\frac{\mu(\lambda x)}{x^{5}}\int_{\Sigma_{C}}dY_{c}\frac{e^{i\lambda x\tilde{\Theta}(Y_{C})}}{\sqrt{\det H_{r}(Y_{C})}}
\end{equation}
where, with $x$ and $\vec{n}_{ef}$ fixed, the only integrations
remaining are over group elements and the phases $\alpha_{ef}$, and
the face amplitude $\mu$ becomes no more than a pre-factor. The critical
surface $\Sigma_{C}$ in this new expression is $S^{2}\times U(1)^{3}$,
corresponding to the one free vector $\vec{n}\in S^{2}$ and the three
free phases $\alpha_{AB},\,\alpha_{BC},\,\alpha_{CA}$ necessary to
define the respective coherent states.

\subsection{Geometric interpretation}

\noindent We will attempt to find a compact expression for the deficit
angle $\tilde{\Theta}$ using the new data. The ``quantum deficit
angle'' for $\Delta_{3}$ is\\
\emph{\scriptsize{
\begin{eqnarray}
\tilde{\Theta} & = & \pm2i\sum_{\pm}(1\pm\gamma)\left[\log\left\langle \mathcal{J}n_{CA}\right|\left(g_{C4}^{\pm}\right)^{\dagger}g_{C5}^{\pm}\left|n_{CB}\right\rangle +\log\left\langle \mathcal{J}n_{BC}\right|\left(g_{B6}^{\pm}\right)^{\dagger}g_{B4}^{\pm}\left|n_{BA}\right\rangle +\log\left\langle \mathcal{J}n_{AB}\right|\left(g_{A5}^{\pm}\right)^{\dagger}g_{A6}^{\pm}\left|n_{AC}\right\rangle \right]\nonumber \\
 & = & \pm2i\sum_{\pm}(1\pm\gamma)\left[\log\left\langle n_{AC}\right|\left(g_{C4}^{\pm}\right)^{\dagger}g_{C5}^{\pm}\left|n_{CB}\right\rangle +\log\left\langle n_{CB}\right|\left(g_{B6}^{\pm}\right)^{\dagger}g_{B4}^{\pm}\left|n_{BA}\right\rangle +\log\left\langle n_{BA}\right|\left(g_{A5}^{\pm}\right)^{\dagger}g_{A6}^{\pm}\left|n_{AC}\right\rangle \right]\label{eq:angleformula}
\end{eqnarray}
}}We will focus on the first of the three matrix elements in the above
expression. The results for the other two can be easily extrapolated
by symmetry. In order to perform the necessary computations, we will
use the following parameterizations of $SU(2)$ and the Hilbert space
$\mathcal{H}^{1/2}$ of spin $\frac{1}{2}$ states:
\begin{itemize}
\item For the SU(2) variables, we use the decomposition
\begin{equation}
\forall g\in\text{SU(2)},\, g=z^{\alpha}\Sigma_{\alpha},\,\left(z^{0}\right)^{2}+\left(z^{1}\right)^{2}+\left(z^{2}\right)^{2}+\left(z^{3}\right)^{2}=1
\end{equation}
where $\Sigma_{0}=\mathbf{1}$ and $\Sigma_{i}=i\sigma_{i}$ for $i=1,2,3$
($\sigma_{i}$ are the Pauli matrices). SU(2) is therefore diffeomorphic
to $S^{3}$, and considering the change of variables
\begin{eqnarray}
z^{0} & = & \cos\gamma\cos\beta^{1}\nonumber \\
z^{3} & = & \cos\gamma\sin\beta^{1}\nonumber \\
z^{1} & = & \sin\gamma\cos\beta^{2}\nonumber \\
z^{2} & = & \sin\gamma\sin\beta^{2},
\end{eqnarray}
with Jacobian $\frac{\sin(2\gamma)}{2}$, where $0<\beta^{i}<2\pi$
and $0<\gamma<\frac{\pi}{2}$, it follows that a general SU(2) matrix
can be written as
\begin{equation}
g=\left[\begin{array}{cc}
\cos\gamma e^{i\beta^{1}} & i\sin\gamma e^{-i\beta^{2}}\\
i\sin\gamma e^{i\beta^{2}} & \cos\gamma e^{-i\beta^{1}}
\end{array}\right].
\end{equation}

\item For the $\mathcal{H}^{1/2}$ variables, naively, one could parametrize
them as follows:
\begin{equation}
\forall\left|n\right\rangle \in\mathcal{H}^{1/2},\,\left|n\right\rangle =\left[\begin{array}{c}
w^{0}+iw^{1}\\
w^{2}+iw^{3}
\end{array}\right],\left(w^{0}\right)^{2}+\left(w^{1}\right)^{2}+\left(w^{2}\right)^{2}+\left(w^{3}\right)^{2}=1
\end{equation}
obtaining $\int_{\mathcal{H}^{1/2}}dn=\int_{S^{3}}dw$. However, it
is advantageous to consider a change of variables that reflects the
construction of a coherent state. Recall that
\begin{equation}
\left|n\right\rangle =e^{i\alpha}G(\vec{n})\left|+\right\rangle 
\end{equation}
where $\vec{n}\in S^{2}$, $\alpha$ is an undetermined phase and
$\left|+\right\rangle =(1,0)$ is the eigenstate of $J_{z}$ with
eigenvalue $+\frac{1}{2}$. The SU(2) element $G(\vec{n})$ is the
rotation that takes $\vec{z}$ to $\vec{n}$ and is readily calculated.
Consider the parameterization of $S^{2}$ in spherical coordinates
\begin{equation}
\vec{n}=(\sin\theta\cos\phi,\sin\theta\sin\phi,\cos\theta)
\end{equation}
To go from $\vec{z}$ to $\vec{n}$ we perform a rotation of angle
$\theta$ around the axis $\vec{n}_{\perp}=(-\sin\phi,\cos\phi,0)$.
From this we get
\begin{eqnarray}
G(\vec{n}) & = & \exp\left(\frac{i\theta}{2}\vec{\sigma}\cdot\vec{n}_{\perp}\right)\nonumber \\
 & = & \exp\left(\frac{i\theta}{2}\left(\cos\phi\sigma_{y}-\sin\phi\sigma_{x}\right)\right)\nonumber \\
 & = & \left[\begin{array}{cc}
\cos\frac{\theta}{2} & e^{-i\phi}\sin\frac{\theta}{2}\\
-e^{i\phi}\sin\frac{\theta}{2} & \cos\frac{\theta}{2}
\end{array}\right].\label{eq:ncoord}
\end{eqnarray}
and therefore
\begin{equation}
\left|n\right\rangle =e^{i\alpha}\left[\begin{array}{c}
\cos\frac{\theta}{2}\\
-e^{i\phi}\sin\frac{\theta}{2}
\end{array}\right].
\end{equation}
The Jacobian of the change of coordinates from $\vec{w}$ to $(\theta,\phi,\alpha)$
is $\frac{\sin(\theta)}{2}$.
\end{itemize}
Since the matrix element $\left\langle n_{AC}\right|\left(g_{C4}^{\pm}\right)^{\dagger}g_{C5}^{\pm}\left|n_{CB}\right\rangle $
is a scalar, it does not depend on the choice of basis in $\mathcal{H}^{1/2}$.
Since the vector part for each of the coherent states present is the
same, we will choose a basis in which $\vec{n}_{AB}=\vec{n}=(0,0,1)$
to carry out computations%
\footnote{There appears to be an ambiguity with this choice, coming from the
parameterization of $S^{2}$ in spherical coordinates - $\vec{n}=(0,0,1)$
is obtained when $\theta=0$, which makes $\phi$ undefined. But it
is evident from (\ref{eq:ncoord}) that $G(0,0,1)=\boldsymbol{1}$.%
}. This translates to
\begin{equation}
\left|n_{i}\right\rangle =e^{i\alpha_{i}}\left[\begin{array}{c}
1\\
0
\end{array}\right],
\end{equation}
for $i\in\{BA,CB,AC\}$. Notice that due to each of the coherent states
appearing exactly once as a bra and a ket in (\ref{eq:angleepsilon}),
the contribution of the phases $\alpha_{i}$ will cancel out and we
can just consider $\left|n\right\rangle =\left[\begin{array}{c}
1\\
0
\end{array}\right]$ from now on. With the coherent states taken care of, we can move
on to $g_{C4}^{\pm}$ and $g_{C5}^{\pm}$. We need to use the gluing
conditions (\ref{eq:gluing}) to relate the two in order to exhaust
the constraints incurring from them, so we will also need an expression
for $R(g)$ for $g\in SU(2)$. Westra%
\footnote{http://www.mat.univie.ac.at/\textasciitilde{}westra/so3su2.pdf%
} gives us a parameterization for $g=\left[\begin{array}{cc}
x & y\\
-\bar{y} & \bar{x}
\end{array}\right]$, $\left|x\right|^{2}+\left|y\right|^{2}=1$:
\begin{equation}
R(g)=\left[\begin{array}{ccc}
\Re(x^{2}-y^{2}) & \Im(x^{2}+y^{2}) & -2\Re(xy)\\
-\Im(x^{2}-y^{2}) & \Re(x^{2}+y^{2}) & 2\Im(xy)\\
2\Re(x\bar{y}) & 2\Im(x\bar{y}) & \left|x\right|^{2}-\left|y\right|^{2}
\end{array}\right]
\end{equation}
In our set of coordinates for SU(2), $x=\cos\gamma e^{i\beta^{1}}$
and $y=i\sin\gamma e^{-i\beta^{2}}$, hence we can write
\begin{equation}
R(g)=\left[\begin{array}{ccc}
\cos^{2}\gamma\cos(2\beta^{1})+\sin^{2}\gamma\cos(2\beta^{2}) & \cos^{2}\gamma\sin(2\beta^{1})+\sin^{2}\gamma\sin(2\beta^{2}) & \sin(2\gamma)\sin(\beta^{1}-\beta^{2})\\
-\cos^{2}\gamma\sin(2\beta^{1})+\sin^{2}\gamma\sin(2\beta^{2}) & \cos^{2}\gamma\cos(2\beta^{1})-\sin^{2}\gamma\cos(2\beta^{2}) & \sin(2\gamma)\cos(\beta^{1}-\beta^{2})\\
\sin(2\gamma)\sin(\beta^{1}+\beta^{2}) & -\sin(2\gamma)\cos(\beta^{1}+\beta^{2}) & \cos(2\gamma)
\end{array}\right]\label{eq:mydoublecover}
\end{equation}
While daunting at first, this expression becomes more tractable within
the context of the gluing condition and the basis choice we made for
$\vec{n}_{AB}$. The gluing condition is reduced to
\begin{equation}
\left[\begin{array}{c}
\sin(2\gamma_{A})\sin(\beta_{A}^{1}-\beta_{A}^{2})\\
\sin(2\gamma_{A})\cos(\beta_{A}^{1}-\beta_{A}^{2})\\
\cos(2\gamma_{A})
\end{array}\right]=\left[\begin{array}{c}
\sin(2\gamma_{B})\sin(\beta_{B}^{1}-\beta_{B}^{2})\\
\sin(2\gamma_{B})\cos(\beta_{B}^{1}-\beta_{B}^{2})\\
\cos(2\gamma_{B})
\end{array}\right]
\end{equation}
where the variables labelled $A$ pertain to $g_{A2}$ and the ones
labelled $B$ pertain to $g_{B1}$, and we omit the $\pm$ index for
simplicity. It is clear that the gluing condition does not fix $g_{A2}$
completely given $g_{B1}$, since they only depend on the differences
$\beta_{A,B}^{1}-\beta_{A,B}^{2}\equiv\delta_{A,B}$. Analysing the
equations,
\begin{itemize}
\item the third equation implies $\gamma_{A}=\gamma_{B}=\gamma$, since
$2\gamma_{A,B}\in[0,\pi]$ and the cosine function is injective in
this domain;
\item given that $\gamma_{A}=\gamma_{B}$, the first and second equations
read $\sin\delta_{A}=\sin\delta_{B}$ and $\cos\delta_{A}=\cos\delta_{B}$,
which for $\delta_{A,B}\in[0,2\pi]$ is enough to infer $\delta_{A}=\delta_{B}$.
\end{itemize}
Hence, we have that, in our chosen basis for $\mathcal{H}^{1/2}$,
if $g_{C4}^{\pm}$ is given by the coordinates $(\gamma_{\pm},\beta_{\pm}^{1},\beta_{\pm}^{2})$,
then $g_{C5}^{\pm}$ is given by $(\gamma_{\pm},\beta_{\pm}^{1}+\epsilon_{\pm},\beta_{\pm}^{2}+\epsilon_{\pm})$
where $\epsilon_{\pm}\in[0,2\pi[$. We can now compute $\left\langle n\right|\left(g_{C4}^{\pm}\right)^{\dagger}g_{C5}^{\pm}\left|n\right\rangle $:\emph{\small{
\begin{eqnarray}
\left\langle n\right|\left(g_{C4}^{\pm}\right)^{\dagger}g_{C5}^{\pm}\left|n\right\rangle  & = & \left[\begin{array}{cc}
1 & 0\end{array}\right]\left[\begin{array}{cc}
\cos\gamma e^{-i\beta^{1}} & -i\sin\gamma e^{-i\beta^{2}}\\
-i\sin\gamma e^{i\beta^{2}} & \cos\gamma e^{i\beta^{1}}
\end{array}\right]\left[\begin{array}{cc}
\cos\gamma e^{i(\beta^{1}+\epsilon)} & i\sin\gamma e^{-i(\beta^{2}+\epsilon)}\\
i\sin\gamma e^{i(\beta^{2}+\epsilon)} & \cos\gamma e^{-i(\beta^{1}+\epsilon)}
\end{array}\right]\left[\begin{array}{c}
1\\
0
\end{array}\right].\nonumber \\
 & = & \left[\begin{array}{cc}
\cos\gamma e^{-i\beta^{1}} & \sin\gamma e^{-i\beta^{2}}\end{array}\right]\left[\begin{array}{c}
\cos\gamma e^{i(\beta^{1}+\epsilon)}\\
\sin\gamma e^{i(\beta^{2}+\epsilon)}
\end{array}\right]\nonumber \\
 & = & e^{i\epsilon}
\end{eqnarray}
}}Taking logarithms, we get simply $i\epsilon$, and substituting
(with proper labels) on the expression for $\tilde{\Theta}$ and repeating
the process for the other two inner products in $\tilde{\Theta}$
(we shall identify the variables pertaining to each of these terms
with an index $i\in\{1,2,3\}$), we obtain
\begin{eqnarray}
\tilde{\Theta} & = & \pm2\sum_{\pm}(1\pm\gamma)\sum_{i=1}^{3}\epsilon_{i}^{\pm}.\label{eq:angleepsilon}
\end{eqnarray}
Remember that all $g_{ve}$ have been determined earlier using the
interior-boundary conditions. Therefore, the $\epsilon_{i}^{\pm}$
can be expressed in terms of the boundary data through some simple
algebra. We give an example. $R(g_{A5}^{\pm})$ and $R(g_{A6}^{\pm})$
are known. Let's call them $A,\, B$ for simplicity. Using the parameterization
(\ref{eq:mydoublecover}), we want to find either $\beta^{1}$ or
$\beta^{2}$ for each matrix, and take their difference to obtain
$\epsilon$. Step by step:
\begin{itemize}
\item $\gamma$ is obtained through $\cos(2\gamma)=A_{33}$. Since $2\gamma\in[0,\pi]$,
the cosine function is injective in this domain and we can write $\gamma=\frac{1}{2}\cos^{-1}(A_{33}).$
There will be three cases to consider due to the possibility of $\sin(2\gamma)$
being zero.
\item If $0<\gamma<\pi/2$, it's easy to extract the sine and cosine of
$\beta^{1}\pm\beta^{2}$ through $A_{31},\, A_{32}$ and $A_{12},\, A_{13}$
respectively. The angles can then be obtained using the angle function
$\mathcal{A}_{1}(x,y)\equiv2\tan^{-1}\left(\frac{x}{1+y}\right)$.
The result for $\beta^{1}$ is
\begin{equation}
\beta^{1}=\frac{1}{2}\left[\mathcal{A}_{1}\left(\frac{A_{13}}{\sqrt{1-A_{33}^{2}}},\frac{A_{23}}{\sqrt{1-A_{33}^{2}}}\right)+\mathcal{A}_{1}\left(\frac{A_{31}}{\sqrt{1-A_{33}^{2}}},\frac{A_{32}}{\sqrt{1-A_{33}^{2}}}\right)\right]
\end{equation}

\item If $\gamma=0$, it is readily seen that $R(g)$ does not depend on
$\beta^{2}$ but $\beta^{1}$ has a simple expression
\begin{equation}
\beta^{1}=\frac{1}{2}\mathcal{A}_{1}\left(A_{12},\, A_{11}\right)
\end{equation}

\item If $\gamma=\pi/2$, $R(g)$ does not depend on $\beta^{1}$ instead.
$\beta^{2}$ is found to be
\begin{equation}
\beta^{2}=\frac{1}{2}\mathcal{A}_{1}\left(A_{12},\, A_{11}\right)
\end{equation}
so we can combine the two extremal cases into one, as they give the
same formal expression for $\epsilon$.
\end{itemize}
Why the emphasis on determining the $\epsilon_{i}^{\pm}$? As seen
in (\ref{eq:angleepsilon}), the deficit angle $\tilde{\Theta}$ has
a very simple expression in terms of them, and they can be interpreted
geometrically. Indeed, note that the expression for $\tilde{\Theta}_{f}$
in a general face can be written as a sum over vertices, $\tilde{\Theta}_{f}=\sum_{v\in f}\tilde{\Theta}_{vf}$.
We know from Han/Zhang's work (among others) that the action is interpreted
as a holonomy around a certain face, going through all the vertices
it belongs to. And in the expression for $\tilde{\Theta}_{vf}$, 
\begin{eqnarray}
\tilde{\Theta}_{vf} & = & \sum_{\pm}2(1\pm\gamma)\log\left\langle \mathcal{J}n_{ef}\right|g_{ev}^{\pm}g_{ve'}^{\pm}\left|n_{e'f}\right\rangle \\
 & \sim & \sum_{\pm}2(1\pm\gamma)\epsilon_{i}^{\pm}
\end{eqnarray}
the inner product clearly illustrates the parallel transport between
the two tetrahedra in $v$ which contain $f$. Therefore $\tilde{\Theta}_{vf}$
can be associated to the internal angle $\angle(e,e')_{vf}$, as illustrated
by the figure below, a two dimensional sketch of the geometric structure
around a vertex.

\begin{center}
\includegraphics[scale=0.75]{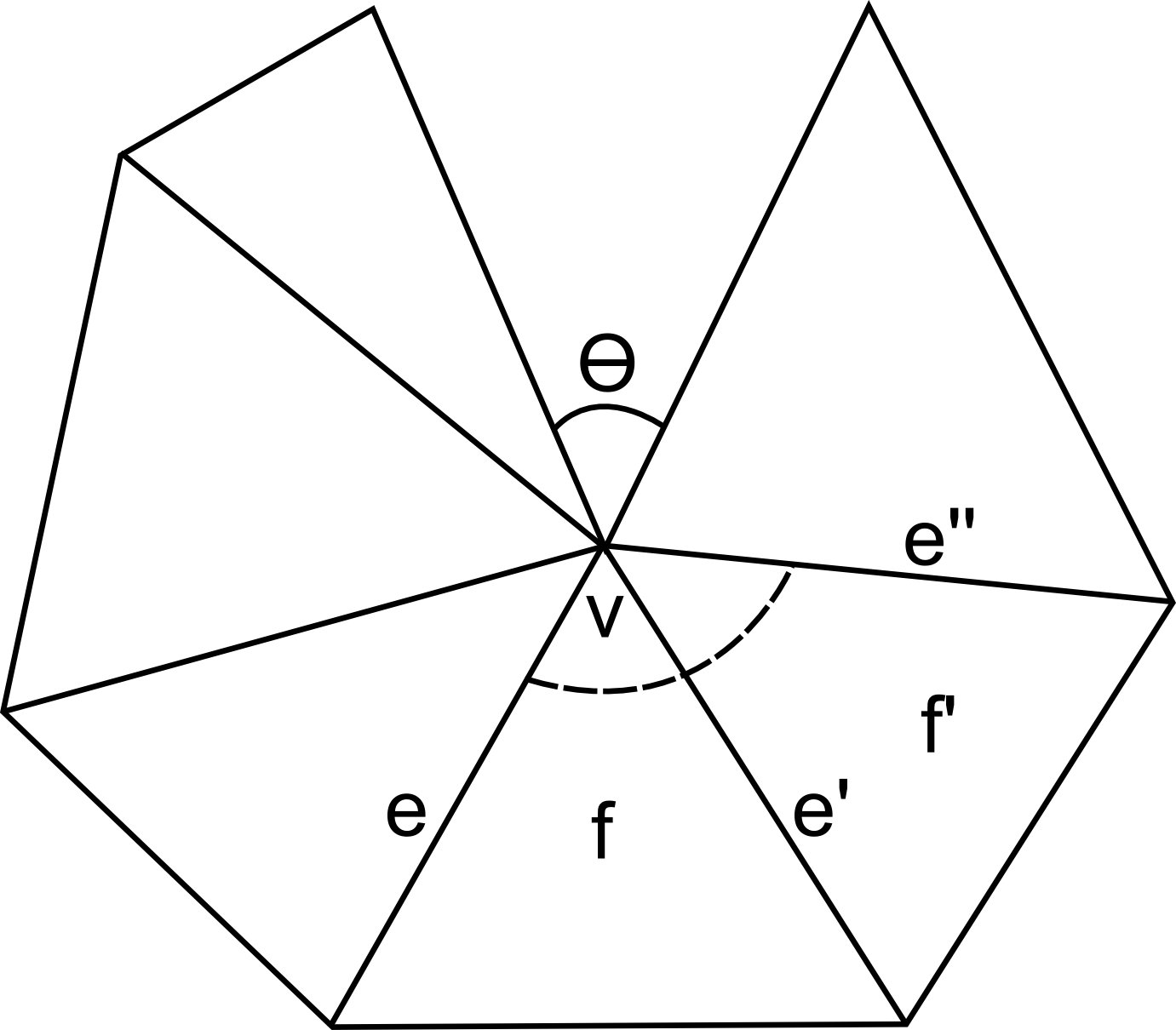}
\par\end{center}

\noindent The sum of all internal angles is equal to $2\pi$ minus
the deficit angle $\Theta_{Regge}$, while the sum of all the $\tilde{\Theta}_{vf}$
should tend asymptotically to a sign factor times $i\gamma\Theta_{Regge}$.
Hence, the correct identification which relates the $\epsilon$ to
the internal angles is
\begin{equation}
\pm\frac{i}{\gamma}\tilde{\Theta}_{vf}=\pm\frac{2i}{\gamma}\sum_{\pm}\sum_{i}(1\pm\gamma)\epsilon_{i}^{\pm}\sim\angle(e,e')_{vf}
\end{equation}
The results obtained in this section seem positive towards the consistency
of EPRL/FK asymptotics with Regge calculus, in contradiction with
the flatness problem, since we are able to obtain geometrically consistent
values for the key quantities in this problem, the area $\gamma j$
and the deficit angle $\Theta$ of the only interior triangle in the
manifold. In fact, a similar result has been claimed by Perini and
Magliaro\cite{key-100}, although the paper in question does not treat
the problem in detail and fails to address one important difficulty
which we will now mention: the behaviour of the state contributions
when $j$ is varied. This is a problem because $j$ is discrete, and
while we get equations of motion that guarantee the nonexistence of
a critical point when $j$ is different from the unique value $j_{0}$
found above, it has not been properly justified that the contribution
from this point is dominant over certain non-critical configurations
with different values of $j$, since it is unclear how to vary the
action over it. Additionally, the value of $j$ that solves exactly
the closure conditions will in general be a non-integer, therefore
there is some uncertainty in this calculation which is important to
address. The closure conditions will, in general, not be exactly satisfied,
because of the discreteness feature.

\subsection{Variation over $j$}

To address the issue, we will use results from Chapter 7 of \cite{key-97}
related to the stationary phase method. In particular we are interested
in the following theorem about the study of the stationary phase integral
when the functions that define it depend on free parameters.\\
\\
\uline{Theorem:}\emph{ Let $f(x,y)$ be a complex valued $C^{\infty}$
function in a neighbourhood $K$ of $(0,0)\in\mathbb{R}^{n+m}$, such
that $\Im(f)\geq0$, $\Im(f(0,0))=0$, $D_{x}f(0,0)=0$ and $\det\, D_{x}^{2}f(0,0)\neq0$.
Let $u$ be a $C^{\infty}$ function with compact support in $K$.
Then
\begin{equation}
\int u(x,y)e^{i\lambda f(x,y)}dx\underset{\lambda\rightarrow\infty}{\sim}e^{i\lambda f^{0}}\left(\frac{2\pi i}{\lambda}\right)^{n/2}\sqrt{\frac{1}{\det D_{x}^{2}f(0,y)^{0}}}\label{eq:theorem}
\end{equation}
where the superscript 0 in front of the determinant signals that the
corresponding function is specified modulo the ideal I of functions
generated by the derivatives $D_{x}f(x,y)$.}\\
\\
Essentially, what the theorem states is that if $x=0$ is a critical
point of $f$ when the free parameter $y$ is zero, then when $y$
is non-zero the point is ``moved'', and is in general not a critical
point any more, but its contribution to the full integral is approximated
by the formula above. The key point is that if $f^{0}$ has an imaginary
part, this contribution is suppressed by a factor $e^{-\lambda\Im(f^{0})}$.
We are interested in this suppression factor for the integral we are
studying, where the free parameter $y$ is taken to be $x-x_{0}$,
$x_{0}$ being the critical value of $x$. But what is $f^{0}$? The
proof of the theorem above uses the Malgrange preparation theorem,
also explained in Chapter 7 of \cite{key-97}. Basically, one can
choose a set of functions $X^{i}(y)$ satisfying $X^{i}(0)=0$ such
that the ideal $I$ of functions generated by $\frac{\partial f}{\partial x^{i}}$
is also generated by $\{x^{i}-X^{i}(y)\}_{i}$, and using the Malgrange
preparation theorem it is possible to write the following expansion
for $f(x,y)$ near the critical point:
\begin{equation}
f(x,y)\approx\sum_{\left|\alpha\right|<N}\frac{f^{\alpha}(y)}{\alpha!}(x-X(y))^{\alpha}\,\text{mod}\, I^{N},\,\forall N
\end{equation}
$f^{0}$ is the term independent of $x$ in this expansion. It is
also noted that the $f_{i}^{1}(y)$ belong to $I^{N}$ for any $N$,
so that they can be chosen to vanish - which is an intuitive result
when compared to a Taylor expansion around a critical point. Since
we are only looking for the leading term of $f^{0}$ to be able to
obtain the suppression factor, we will consider an expansion to second
order $(N=2)$, and to compute the different functions in play we
will use the well-known Taylor series for $f$: 
\begin{eqnarray}
f(x,y) & \approx & f(0,0)+\underbrace{\left.\frac{\partial f}{\partial x^{i}}\right|_{(0,0)}}_{=0}x^{i}+\underbrace{\left.\frac{\partial f}{\partial y}\right|_{(0,0)}}_{\equiv\delta_{1}}y\\
 & + & \frac{1}{2}\underbrace{\left.\frac{\partial^{2}f}{\partial y\partial x^{i}}\right|_{(0,0)}}_{\equiv K_{i}}yx^{i}+\frac{1}{2}\underbrace{\left.\frac{\partial^{2}f}{\partial y^{2}}\right|_{(0,0)}}_{\equiv\delta_{2}}y^{2}+\frac{1}{2}\underbrace{\left.\frac{\partial^{2}f}{\partial x^{i}\partial x^{j}}\right|_{(0,0)}}_{\equiv H_{ij}}x^{i}x^{j}
\end{eqnarray}
The second order Malgrange expansion for $f(x,y)$ is (setting $f^{1}=0$)
\begin{equation}
f(x,y)\approx f^{0}(y)+\frac{1}{2}f_{ij}^{2}(y)(x^{i}-X^{i}(y))(x^{j}-X^{j}(y))
\end{equation}
Equating both expansions and gathering terms independent, linear and
quadratic in $x$, we get
\begin{eqnarray}
f(0,0)+\delta_{1}y+\frac{1}{2}\delta_{2}y^{2} & = & f^{0}+\frac{1}{2}f_{ij}^{2}X^{i}X^{j}\nonumber \\
\frac{1}{2}K_{i}x^{i}y & = & -\frac{1}{2}\left(f_{ij}^{2}+f_{ji}^{2}\right)x^{i}X^{j}\nonumber \\
\frac{1}{2}H_{ij}x^{i}x^{j} & = & \frac{1}{2}f_{ij}^{2}x^{i}x^{j}
\end{eqnarray}
which we solve to obtain ($H^{ij}$ is the inverse matrix of $H_{ij}$.
Remember we assumed $\det H\neq0$) 
\begin{eqnarray}
f^{0} & = & f(0,0)+\delta_{1}y+\frac{1}{2}\delta_{2}y^{2}-\frac{1}{2}K_{i}H^{ij}K_{j}y^{2}\nonumber \\
-H^{ij}K_{i}y & = & X^{j}\nonumber \\
f_{ij}^{2} & = & H_{ij}
\end{eqnarray}
Applying to the $\Delta_{3}$ case, remembering that we chose $y=x-x_{0}$,
we see that $f(0,0)$ is the action at the critical point $S_{C}$,
$\delta_{1}=-i\tilde{\Theta}_{C}\sim\pm\gamma\Theta_{Regge}$ and
$\delta_{2}=0$. Note that $\delta_{1}$ is real. We are only interested
in the imaginary part of $f^{0}$, which is quadratic in $(x-x_{0})$,
and gives us the suppressing factor as
\begin{equation}
\exp\left(\frac{\lambda}{2}\Im\left(K_{i}H^{ij}K_{j}\right)(x-x_{0})^{2}\right)\label{eq:dagauss}
\end{equation}
Note that the variation of $x$ has to be discrete. We would set $j=j_{0}+\frac{n}{2},\, n\in\mathbb{Z}$,
so that $x-x_{0}=\frac{n}{2\lambda}$. This allows us to write the
partition function as a sum over $n$ in terms of the term corresponding
to $n=0$, the critical term:
\begin{equation}
Z=Z_{C}\sum_{n}\exp\left(-\frac{A}{4\lambda}n^{2}\right)\label{eq:gaussseries}
\end{equation}
where $A=-\Im\left(K_{i}H^{ij}K_{j}\right)$. If $x$ is thought of
as an approximately continuous variable, the distribution of $x$
values follows a Gaussian curve with standard deviation $\sigma=\sqrt{\frac{1}{\lambda A}}$.
This is a sufficiently small deviation, assuming $A$ finite, to conclude
that the distribution of the $(j_{f},g_{ve},n_{ef})$ variables is
sufficiently peaked around the critical surface. Since $A$ does not
have any $\lambda$ dependence, the positive result should be guaranteed
simply by $A\neq0$. However, the most rigorous approach to this problem
is to compute the sum of the series in (\ref{eq:gaussseries}) and
obtain the statistics of the discrete variable $n$ (note, in particular,
that $j_{0}$ as given by the closure equations might not be a semi-integer,
so the dominant contribution would come from the semi-integer closest
to it). The EPRL/FK action 
\begin{equation}
S=-2i\sum_{f}\sum_{v\in f}\sum_{\pm}j_{f}(1\pm\gamma)\log\left\langle \mathcal{J}n_{ef}\right|\left(g_{ve}^{\pm}\right)^{\dagger}g_{ve'}^{\pm}\left|n_{e'f}\right\rangle 
\end{equation}
can be interpreted in terms of this stationary phase method by setting
$j_{f}\equiv y$ as the free parameter, and $x_{i}\equiv\left(\left\{ g_{ve}\right\} _{a},\left\{ n_{ef}\right\} _{b}\right)$
as the dependent variables, where $a,\, b$ signal an appropriate
coordinate system in which to express the interior $g_{ve},\, n_{ef}$
(which can be, for example, the parameterizations of SU(2) and $\mathcal{H}^{1/2}$
specified in section 4.2). The quantities necessary to compute the
approximate partition function (\ref{eq:gaussseries}) are
\begin{eqnarray}
K_{i} & = & \left.\frac{\partial^{2}S}{\partial j_{f}\partial x_{i}}\right|_{\text{critical}}=\left.\frac{\partial\tilde{\Theta}_{f}}{\partial x_{i}}\right|_{\text{critical}}\label{eq:derivativeswewant}\\
H_{ij} & = & \left.\frac{\partial^{2}S}{\partial x_{i}\partial x_{j}}\right|_{\text{critical}}
\end{eqnarray}
where ``critical'' means the derivatives are computed at the unique
critical point for $\Delta_{3}$ determined in section 4.1, and $K_{i}$
is simplified due to the action being linear in $j$, being reduced
to first derivatives of the quantum deficit angle of the interior
face $\tilde{\Theta}_{f}$. The conditions of theorem (\ref{eq:theorem})
require that $\det\, H\neq0$ for the stationary phase method to be
applicable. However, explicit computation of this determinant, even
using algebraic computation software, proves to be a bit too cumbersome
because of the dependence of the derivatives in question on a high
number of \emph{a priori} arbitrary boundary variables, $\left\{ g_{ve},\, n_{ef}\right\} _{B}$
- even though it is possible to compute $det\, H$ explicitly in terms
of them, and obtain a numeric answer if numeric data are introduced
for the EPRL variables, it is not clear at the moment whether, for
example, it is nonzero for all their possible values. For that reason,
we will analyse the determination of EPRL boundary data from geometric
constructions, in order to obtain values for $H$ in concrete cases.
\\
\\
While showing consistency of the EPRL behaviour with Einstein theory
in such examples is in no way a proof for the general case even within
$\Delta_{3}$, it would nevertheless be an interesting result, and
on the flipside, an inconsistency would be a significant result on
its own, albeit a negative one. To summarize the possible outcomes:
\begin{itemize}
\item $det\, H=0$: then the stationary phase method is not valid (in particular
the quantity $A$ is not defined), and we must find a different method
to evaluate the asymptotics;
\item $det\, H\neq0$ and $A=0$: in that case the Gaussian distribution
(\ref{eq:gaussseries}) has infinite standard deviation and as such
will not specify the semiclassical value of $x$, failing to reproduce
the expected classical result;
\item $det\, H\neq0$ and $A\neq0$: the Gaussian distribution around the
semiclassical value of $x$ should guarantee reproduction of the expected
geometric values. In particular, if one can verify this to happen
for a certain boundary configuration, continuity conditions assure
that the EPRL asymptotics match the expected classic solutions in
a certain open neighbourhood of that configuration, which would give
us some confidence that the semiclassical limit is correct for a significant
range of boundary data. It does not, however, discard the possibility
of there existing isolated points in the critical surface for which
one of the two situations above happen, and it is unclear how this
would affect the overall statistics.
\end{itemize}

\subsection{Constructing EPRL spin foam variables from geometrical data}

To obtain the EPRL spin foam variables $g_{ve}$, $\vec{n}_{ef}$,
$j_{f}$ for a given example, we need to essentially carry the procedure
of the reconstruction theorem backwards and determine how they are
related to the geometrical data which defines the classical triangulation
$\Delta$. Obtaining the spins $j_{f}$ is straightforward. Indeed,
it has already been established that $j_{f}$ are directly related
to the triangle areas via $A_{f}=\gamma j_{f}$ (within our semiclassical
approximation of $j$ being large). \\
\\
The Livine-Speziale coherent states $\left|n_{ef}\right\rangle $
are expressed in terms of $\vec{n}_{ef}\in S^{2}$, the normal vectors
to the tetrahedron faces' Euclidean images in the tangent spaces $T_{e}\Delta\approx\mathbb{\mathbb{R}}^{3}$,
and phases $\alpha_{ef}\in\text{U(1) }$ which can be consistently
defined by imposing Regge boundary conditions but are of no consequence
to the dynamics of the model, and can therefore safely be ignored.
The one difficulty in correctly identifying the $\vec{n}_{ef}$ is
that computing the norms of the geometrical tetrahedra in $\mathbb{R}^{3}$
does not immediately tell you which $n_{ef}$ is which within a certain
tetrahedron. A solution to this issue is to consider\emph{ gluing
matrices}. Indeed, considering a gluing equation
\begin{equation}
R(g_{ve}^{\pm})\vec{n}_{ef}=-R(g_{ve'}^{\pm})\vec{n}_{e'f}\,,
\end{equation}
notice that the $+$ and $-$ equations contained in it can both be
manipulated to give the value of $\vec{n}_{e'f}$, and therefore
\begin{equation}
\left(R^{-1}(g_{ve'}^{+})R(g_{ve}^{+})-R^{-1}(g_{ve'}^{-})R(g_{ve}^{-})\right)\vec{n}_{ef}=0.\label{eq:gluingmatrix}
\end{equation}
Defining the matrix in brackets as the gluing matrix between two tetrahedra,
$R_{ee'}$, $\vec{n}_{ef}$ must lie in its null space, and furthermore,
if the tetrahedron is non-degenerate (which we are assuming it is)
such null space must have dimension 1. Comparing the resultant null
spaces with the normals of the geometric tetrahedra then gives the
correct answer for $\vec{n}_{ef}$%
\footnote{It is still necessary to consider the geometric tetrahedra with this
procedure since simply solving (\ref{eq:gluingmatrix}) gives the
correct normals up to a minus sign, which must be fixed in accordance
with geometric consistency.%
}.\\
Obtaining the $g_{ve}$ is somewhat less trivial. The first step is
to identify what they represent geometrically. Indeed, $g_{ve}$ are
SO(4) group elements related to the triangulated equivalent of the
spin connection, which in the geometrical setup translates to mapping
the geometrical tetrahedron $e\in v$ to its image in the tangent
space $T_{e}\Delta$. We have to define what this means, though.\\
\\
Consider a 4-simplex $v\in\Delta$ and a tetrahedron $e\in v$ defined
by points $p_{1},...,p_{4}$. Note that for a general triangulation
each 4-simplex lives on its own copy of $\mathbb{R}^{4}$: if the
entire triangulation can be embedded isometrically in $\mathbb{R}^{4}$
that implies all the deficit angles are zero and the triangulation
is flat. We will define the tetrahedron's geometric matrix $M_{ve}$
and projected matrix $M_{ve}^{(3)}$ :
\begin{itemize}
\item to construct $M_{ve}$, consider an oriented trivector $\tau_{ve}=\left\{ \tau_{ve}^{1},\tau_{ve}^{2},\tau_{ve}^{3}\right\} $
consisting of the three edge vectors coming out of a previously defined
pivot point. For example, if $p_{1}$ is chosen as the pivot, a possible
trivector is $\{p_{2}-p_{1},p_{3}-p_{1},p_{4}-p_{1}\}$. If $e$ is
non-degenerate, the trivector defines a (non-orthonormal) basis of
the 3-dimensional hyperplane $e$ lives on, which can be equated to
$T_{e}\Delta$. Compute the normal to this hyperplane, $N_{ve}$,
which is the normal to the tetrahedron. Note that there are two possible
orientations for this normal, so we will establish as a convention
that the orientation to choose is the one that makes $\det\, M_{ve}>0$.
The full matrix is then 
\begin{equation}
M_{ve}=\left\{ N_{ve},\tau_{ve}^{1},\tau_{ve}^{2},\tau_{ve}^{3}\right\} .
\end{equation}
Note that this matrix is, by construction, invertible, since its 4
columns are linearly independent.
\item for $M_{ve}^{(3)}$, write down an orthonormal basis of $T_{e}\Delta$
as defined above, for example using the Gram-Schmidt orthonormalization
algorithm, and determine the coordinates of the vectors in $\tau_{ve}$
on that basis. Call them $\tau_{ve}^{(3)}$. We will regard $T_{e}\Delta$
as a subspace of $\mathbb{R}^{4}$ normal to $(1,0,0,0)$, since it
will help with decomposing $g_{ve}$ into its SU(2) components $g_{ve}^{\pm}$.
The projected tetrahedron matrix is then 
\begin{equation}
M_{ve}^{(3)}=\left[\begin{array}{cccc}
1 & 0 & 0 & 0\\
0\\
0 & \left(\tau_{ve}^{1}\right)^{(3)} & \left(\tau_{ve}^{2}\right)^{(3)} & \left(\tau_{ve}^{3}\right)^{(3)}\\
0
\end{array}\right]
\end{equation}
which is also invertible by the same reasons as above.
\end{itemize}
Note that $M_{ve}$ is not unique to a tetrahedron, but the $g_{ve}$
rotation will be well defined provided that the orientations of both
are consistent with respect to the considerations of section 2, that
is, deriving the orientation of each tetrahedron from the 4-simplex
$v$ by (\ref{eq:orient}) and permuting the edge vectors in $\tau_{ve}$
to guarantee the same sign for all $M_{ve}$ associated with $v$.
With these definitions in place, $g_{ve}$ is the SO(4) matrix that
rotates the projected matrix into the geometric matrix, i.e.
\begin{eqnarray}
g_{ve}\cdot M_{ve}^{(3)} & = & M_{ve}\nonumber \\
\Leftrightarrow g_{ve} & = & M_{ve}\left(M_{ve}^{(3)}\right)^{-1}
\end{eqnarray}
Next step is to find $g_{ve}$'s SU(2) components. To do this we will
use a result of van Elfrinkhof\cite{key-104} which gives an algorithm
for decomposition of a SO(4) rotation into left- and right-isoclinic
rotations, which can each be associated to SU(2) elements. Given a
matrix $g\in$ SO(4), define the associate matrix{\scriptsize{
\begin{equation}
\text{Asc}(g)=\frac{1}{4}\left[\begin{array}{cccc}
g_{00}+g_{11}+g_{22}+g_{33} & g_{10}-g_{01}-g_{32}+g_{23} & g_{20}+g_{31}-g_{02}-g_{13} & g_{30}-g_{21}+g_{12}-g_{03}\\
g_{10}-g_{01}+g_{32}-g_{23} & -g_{00}-g_{11}+g_{22}+g_{33} & g_{30}-g_{21}-g_{12}+g_{03} & -g_{20}-g_{31}-g_{02}-g_{13}\\
g_{20}-g_{31}-g_{02}+g_{13} & -g_{30}-g_{21}-g_{12}-g_{03} & -g_{00}+g_{11}-g_{22}+g_{33} & g_{10}+g_{01}-g_{32}-g_{23}\\
g_{30}+g_{21}-g_{12}-g_{03} & g_{20}-g_{31}+g_{02}-g_{13} & -g_{10}-g_{01}-g_{32}-g_{23} & -g_{00}+g_{11}+g_{22}-g_{33}
\end{array}\right].
\end{equation}
}}van Elfrinkhof's theorem states that $\text{Asc}(g)$ has rank one
and is normalized under the Euclidean norm, $\sum_{ij}\left(\text{Asc}(g)_{ij}\right)^{2}=1$,
and that there exists a duo of vectors $(a,b,c,d)$ and $(p,q,r,s)$
in $S^{3}\times S^{3}$ such that
\begin{equation}
\text{Asc}(g)=\left[\begin{array}{cccc}
ap & aq & ar & as\\
bp & bq & br & bs\\
cp & cq & cr & cs\\
dp & dq & dr & ds
\end{array}\right].\label{eq:associatedec}
\end{equation}
More precisely, there are exactly two vector pairs in $S^{3}\times S^{3}$
that satisfy this, since for a given $\left\{ (a,b,c,d),\,(p,q,r,s)\right\} $,
their opposites $\left\{ (-a,-b,-c,-d),\,(-p,-q,-r,-s)\right\} $
also constitute a solution. Since there is a group isomorphism between
$S^{3}$ and SU(2) given by
\begin{eqnarray}
\phi:\, S^{3} & \rightarrow & \text{SU}(2)\nonumber \\
(a,b,c,d) & \rightarrow & a\mathbf{1}+i\left(b\sigma_{1}+c\sigma_{2}+d\sigma_{3}\right),\label{eq:s3su2}
\end{eqnarray}
where $\sigma_{i}$ are the Pauli matrices and $\mathbf{1}$ is the
identity matrix in SU(2), the aforementioned vector duos are directly
mapped to SU(2) group elements. The decomposition is made explicit
within SO(4) by the formula
\begin{equation}
g=\left[\begin{array}{cccc}
a & -b & -c & -d\\
b & a & -d & c\\
c & d & a & -b\\
d & -c & b & a
\end{array}\right].\left[\begin{array}{cccc}
p & -q & -r & -s\\
q & p & s & -r\\
r & -s & p & q\\
s & r & -q & p
\end{array}\right].\label{eq:so4decomposition}
\end{equation}
where the left and right matrices are left-isoclinic and right-isoclinic,
respectively. (\ref{eq:so4decomposition}) can also be specified neatly
in quaternion notation. Consider the set of quaternions $\mathbb{H\approx\mathbb{R}}^{4}$
with the basis vectors $\mathbf{1},\, I,\, J,\, K$. $\mathbb{\mathbb{H}}$
can also be defined in $\mathbb{C}^{2\times2}$ by extending the domain
of the map $\phi$ in (\ref{eq:s3su2}) to all of $\mathbb{R}^{4}$.
Using the latter formulation, the SU(2)$\times$SU(2) action on a
vector $v\in\mathbb{\mathbb{H}}$ is neatly written as
\begin{equation}
(g^{+},\, g^{-})\cdot v=g^{+}v\left(g^{-}\right)^{-1}
\end{equation}
and translates to the action of the SO(4) matrix with $(g^{+},\, g^{-})$
as its left and right isoclinic components according to the van Elfrinkhof
formula. We will use these results to establish the correspondence
\begin{eqnarray}
g_{ve}^{+} & = & \phi(a,b,c,d)\nonumber \\
g_{ve}^{-} & = & \left[\phi(p,q,r,s)\right]^{-1}.
\end{eqnarray}
Now there is an issue with this definition, which is the ambiguity
between which of the two vector pairs that solve the van Elfrinkhof
theorem to choose for each $g_{ve}$ in order to maintain consistency,
since SU(2)$\times$SU(2) double covers SO(4). We will address this
problem by establishing an algorithm. For notation simplicity write
$M\equiv\text{Asc}(g)$. First analyze cases where $M_{11}\neq0$
(resulting that $a,\, p\neq0)$. Define
\begin{equation}
K=\sqrt{M_{11}^{2}+M_{12}^{2}+M_{13}^{2}+M_{14}^{2}}
\end{equation}
Since, using (\ref{eq:associatedec}),
\begin{equation}
p=\frac{M_{11}}{a};\,\, q=\frac{M_{12}}{a};\,\, r=\frac{M_{13}}{a};\,\, s=\frac{M_{14}}{a}
\end{equation}
and $p^{2}+q^{2}+r^{2}+s^{2}=1$, it follows that $a=\pm\sqrt{M_{11}^{2}+M_{12}^{2}+M_{13}^{2}+M_{14}^{2}}=\pm K$.
For the sake of consistency we will always take the positive root
$a=K$. It is then straightforward to obtain
\begin{eqnarray*}
p & = & \frac{M_{11}}{K};\,\, q=\frac{M_{12}}{K};\,\, r=\frac{M_{13}}{K};\,\, s=\frac{M_{14}}{K}
\end{eqnarray*}
\begin{equation}
a=K;\,\, b=K\frac{M_{21}}{M_{11}};\,\, c=\frac{M_{31}}{M_{11}};\,\, d=\frac{M_{41}}{M_{11}}
\end{equation}
Whenever $M_{11}\neq0$ this algorithm provides a consistent definition
of the $g^{+}$ and $g^{-}$, but when $M_{11}=0$ a similar process
can be carried out by choosing a non-zero entry $M_{ij}$ (it exists
since both parameter vectors are non-zero) and defining
\begin{equation}
K=\sqrt{\sum_{l=1}^{4}M_{il}^{2}}.
\end{equation}
If we use the notation $(a,b,c,d)\equiv(x_{1},x_{2},x_{3},x_{4})$
and $(p,q,r,s)=(y_{1},y_{2},y_{3},y_{4})$ then we can define a solution
for them as follows:
\begin{eqnarray}
x_{i} & = & K\nonumber \\
y_{l} & = & \frac{M_{il}}{K},\, l\in\{1,2,3,4\}\nonumber \\
x_{l} & = & K\frac{M_{lj}}{M_{ij}},l\neq i.
\end{eqnarray}
To finalize this section we will mention the two geometrical examples
considered for this study. Given the circumstances of the flatness
problem, it was deemed appropriate to consider a flat and a non-flat
version of $\Delta_{3}$ in calculations. As mentioned above, a flat
triangulation is easily defined by considering an embedding of it
in $\mathbb{R}^{4}$, but it's somewhat less trivial to define a non-flat
one. For the latter we will consider a figure analogous to a triangulation
of $S^{4}$ by taking an embedding of $\Delta_{3}$ into $\mathbb{R}^{5}$
given by an equilateral 5-simplex centered at the origin. This embedding
is defined by assigning the 6 points of $\Delta_{3}$ into the 6 points
of the 5-simplex. \\
\\
Let us define the equilateral 5-simplex by building it ``from the
ground up'' from an equilateral triangle centered at the origin.
A triangle in $\mathbb{R}^{2}$ with the desired characteristics is
given by 
\begin{equation}
\left\{ A_{2},B_{2},C_{2}\right\} =\left\{ \left(-\frac{1}{2},-\frac{1}{2\sqrt{3}}\right),\left(\frac{1}{2},-\frac{1}{2\sqrt{3}}\right),\left(0,\frac{1}{\sqrt{3}}\right)\right\} .
\end{equation}
Adding the third axis $x^{2}$ we see that if a fourth point is $D_{3}=(0,0,a_{3})$,
then the tetrahedron formed by{\scriptsize{
\begin{equation}
\left\{ A_{3},B_{3},C_{3},D_{3}\right\} =\left\{ \left(-\frac{1}{2},-\frac{1}{2\sqrt{3}},-\frac{a_{3}}{3}\right),\left(\frac{1}{2},-\frac{1}{2\sqrt{3}},-\frac{a_{3}}{3}\right),\left(0,\frac{1}{\sqrt{3}},-\frac{a_{3}}{3}\right),\left(0,0,a_{3}\right)\right\} 
\end{equation}
}}is centered in the origin and $a_{3}$ can be fixed to make it equilateral
by forcing $\overline{C_{3}D_{3}}=1$. (Note that if $O_{3}$ is the
centre of the triangle $A_{3}B_{3}C_{3}$ then $O_{3}D_{3}$ is normal
to said triangle and therefore $\overline{A_{3}D_{3}}=\overline{B_{3}D_{3}}=\overline{C_{3}D_{3}}$.)
Solving that constraint gives $a_{3}=\sqrt{\frac{3}{8}}$.\\
\\
Similarly, we construct a 4-simplex under the same conditions by adding
the axis $x^{3}$, defining the point $E_{4}=(0,0,0,a_{4})$ and considering
the 4-simplex\emph{\scriptsize{
\begin{eqnarray}
\left\{ A_{4},B_{4},C_{4},D_{4},E_{4}\right\}  & = & \left\{ \left(-\frac{1}{2},-\frac{1}{2\sqrt{3}},-\frac{1}{3}\sqrt{\frac{3}{8}},-\frac{a_{4}}{4}\right),\left(\frac{1}{2},-\frac{1}{2\sqrt{3}},-\frac{1}{3}\sqrt{\frac{3}{8}},-\frac{a_{4}}{4}\right),\right.\nonumber \\
 &  & \left(0,\frac{1}{\sqrt{3}},-\frac{1}{3}\sqrt{\frac{3}{8}},-\frac{a_{4}}{4}\right),\left(0,0,\sqrt{\frac{3}{8}},-\frac{a_{4}}{4}\right),\nonumber \\
 &  & \left.\left(0,0,0,a_{4}\right)\right\} .
\end{eqnarray}
}}By analogous argument to what we used for the tetrahedron, this
4-simplex is centered in the origin and will be equilateral if $\overline{D_{4}E_{4}}=1$,
which is solved to give $a_{4}=\sqrt{\frac{2}{5}}$.\\
\\
Finally, we add the axis $x^{4}$, define $F_{5}=(0,0,0,0,a_{5})$
and consider the 5-simplex\emph{\scriptsize{
\begin{eqnarray}
\left\{ A_{5},B_{5},C_{5},D_{5},E_{5},F_{5}\right\}  & = & \left\{ \left(-\frac{1}{2},-\frac{1}{2\sqrt{3}},-\frac{1}{3}\sqrt{\frac{3}{8}},-\frac{1}{4}\sqrt{\frac{2}{5}},-\frac{a_{5}}{5}\right),\left(\frac{1}{2},-\frac{1}{2\sqrt{3}},-\frac{1}{3}\sqrt{\frac{3}{8}},-\frac{1}{4}\sqrt{\frac{2}{5}},-\frac{a_{5}}{5}\right),\right.\nonumber \\
 &  & \left(0,\frac{1}{\sqrt{3}},-\frac{1}{3}\sqrt{\frac{3}{8}},-\frac{1}{4}\sqrt{\frac{2}{5}},-\frac{a_{5}}{5}\right),\left(0,0,\sqrt{\frac{3}{8}},-\frac{1}{4}\sqrt{\frac{2}{5}},-\frac{a_{5}}{5}\right),\nonumber \\
 &  & \left.\left(0,0,0,\sqrt{\frac{2}{5}},-\frac{a_{5}}{5}\right),(0,0,0,0,a_{5})\right\} .
\end{eqnarray}
}}The 5-simplex has the characteristics we need if $\overline{E_{5}F_{5}}=1$,
which is satisfied when $a_{5}=\sqrt{\frac{5}{12}}$. The coordinates
of the equilateral 5-simplex to be used are therefore\emph{\scriptsize{
\begin{eqnarray}
\left\{ A_{5},B_{5},C_{5},D_{5},E_{5},F_{5}\right\}  & = & \left\{ \left(-\frac{1}{2},-\frac{1}{2\sqrt{3}},-\frac{1}{3}\sqrt{\frac{3}{8}},-\frac{1}{4}\sqrt{\frac{2}{5}},-\frac{1}{5}\sqrt{\frac{5}{12}}\right),\left(\frac{1}{2},-\frac{1}{2\sqrt{3}},-\frac{1}{3}\sqrt{\frac{3}{8}},-\frac{1}{4}\sqrt{\frac{2}{5}},-\frac{1}{5}\sqrt{\frac{5}{12}}\right),\right.\nonumber \\
 &  & \left(0,\frac{1}{\sqrt{3}},-\frac{1}{3}\sqrt{\frac{3}{8}},-\frac{1}{4}\sqrt{\frac{2}{5}},-\frac{1}{5}\sqrt{\frac{5}{12}}\right),\left(0,0,\sqrt{\frac{3}{8}},-\frac{1}{4}\sqrt{\frac{2}{5}},-\frac{1}{5}\sqrt{\frac{5}{12}}\right),\nonumber \\
 &  & \left.\left(0,0,0,\sqrt{\frac{2}{5}},-\frac{1}{5}\sqrt{\frac{5}{12}}\right),\left(0,0,0,0,\sqrt{\frac{5}{12}}\right)\right\} .
\end{eqnarray}
}}This example is particularly simple in numeric terms since the construction
implies that all triangles have the same area $A_{f}=\sqrt{3}/4$,
and the normal vectors $\vec{n}_{ef}$ can all be derived from the
same equilateral tetrahedron in $\mathbb{R}^{3}$, only taking care
to match their orientations correctly.\\
\\
For the flat example, we considered an embedding of $\Delta_{3}$
in $\mathbb{R}^{4}$ using the coordinates
\begin{eqnarray}
a & = & \left(-\frac{1}{2},-\frac{1}{2\sqrt{3}},0,0\right)\nonumber \\
b & = & \left(\frac{1}{2},-\frac{1}{2\sqrt{3}},0,0\right)\nonumber \\
c & = & \left(0,\frac{1}{\sqrt{3}},0,0\right)\nonumber \\
d & = & \left(0,0,-\frac{1}{2},-\frac{1}{2\sqrt{3}}\right)\nonumber \\
e & = & \left(0,0,\frac{1}{2},-\frac{1}{2\sqrt{3}}\right)\nonumber \\
f & = & \left(0,0,0,\frac{1}{\sqrt{3}}\right).
\end{eqnarray}
The ancillary files annexed to this paper include detailed Mathematica
code for computing the spin foam variables $g_{ve}$, $\vec{n}_{ef}$,
$j_{f}$ of both geometrical configurations, and then using them to
determine the relevant derivatives $K_{i}$ and $H_{ij}$, as well
as the decay parameter $A=-\Im\left(K_{i}H^{ij}K_{j}\right)$. Here
we will only state the results, which unfortunately could only be
obtained in numeric form for the coordinates chosen and a given value
of the Immirzi parameter. Note that the Immirzi parameter must be
consistent with triangle areas to ensure that the values of $j_{f}$
are half-integer, and according to the EPRL prescription $0<\gamma<1$.
The results were
\begin{eqnarray}
\Delta_{3}^{(curved)}: & \text{used}\,\gamma=\frac{\sqrt{3}}{2}, & A=6.62021\nonumber \\
\Delta_{3}^{(flat)}: & \text{used}\,\gamma=\frac{1}{2}\sqrt{\frac{5}{3}}, & A=14.4389
\end{eqnarray}
The significant finding here is that they are both nonzero within
numerical error, and therefore in the examples considered the asymptotic
spin foam analysis matches what is expected from general relativity.

\section{Conclusions and future work}

There are a few remarks that we would like to convey with this work.
The first one is that varying the asymptotic EPRL action with respect
to $j_{f}$, with these being discrete, is a delicate issue, and one
that we do not believe can be tackled by simply ignoring discreteness
and taking some \emph{ad hoc} continuum approximation to be able to
differentiate with respect to those spins. Although that line of thought
was what originally lead to the enunciation of the flatness problem,
Hellmann/Kaminski seem to have recovered it under a more rigorous
approach with their holonomy spin foam formalism. In this work we
attempted to explicitly acknowledge the discreteness of $j$ and study
its effects on the statistics of the partition function, by using
the Malgrange preparation theorem and its corollaries to apply the
stationary phase method, and explicit the distribution with respect
to $j$ in a neighbourhood of the critical point. However, the validity
of this method is dependent on the $A$ quantity defined in section
4.3 being finite and mathematically meaningful, which essentially
comes down to whether the Hessian determinant of the action is non-zero
at the (singular) critical point for any possible boundary configuration.
It is a highly non-trivial task from a computational point of view
to verify this, so for the time being we have settled with finishing
the calculation for the example cases proposed. \\
\\
Indeed, we were able to numerically compute the Hessian of the action
and the quantity $A$ for two example configurations: a curved one
based of an embedding in an equilateral 5-simplex, and a flat one
based of an embedding in Euclidean 4-space. We have found them to
be non-zero for both configurations. This is a positive, albeit incomplete,
sign of consistency of the spin foam model in this example, since
it allows us to assert by continuity arguments that the same is valid
in a neighbourhood of the critical point considered. It would be helpful
to conduct a more detailed statistical analysis of the behaviour of
this example's partition function for values of $j$ near the geometric
one, and that is a question to be considered in subsequent work. Also
interesting would be to gain further insight into the behaviour of
$A$ in different configurations, for example by exploring algebraic
properties of the boundary data such that exact expressions for $A$
could be found in certain subsets of possible configurations - since
obtaining a full expression for all possible boundaries seems too
cumbersome to be feasible.\\
\\
The second remark is the positive result that, for this $\Delta_{3}$
configuration, containing only one interior face whose data are entirely
specified at the classical level by boundary data, it is possible
to recover the expected critical point of the action, corresponding
to the values of area and deficit angle for the interior triangle
that ensure proper geometric gluing. Incidentally, this result also
allows us to perform the converse of the reconstruction theorem and
recover EPRL variables from geometric variables in concrete realizations
of the triangulation. The assertion that the critical point for a
given boundary configuration is unique and corresponds to the expected
classical geometry had already been verified by Perini and Magliaro
in \cite{key-100}, but the subtleties regarding the statistics of
the partition function's distribution over $j$ are not addressed
in their work (it is just assumed that non-critical configurations
are exponentially suppressed), in particular the fact that the classical
$j_{0}$ may not be an integer, and in general the range of $j$ near
$j_{0}$ that contributes significantly to the partition function
(even in the circumstances where stationary phase applies correctly
with $A\neq0$) is dictated by a Gaussian distribution whose width
increases with $\lambda$, although the relative uncertainty $\Delta j/j\approx\Delta j/\lambda$
is suppressed for large $\lambda$. We hope that further analysis
will bring some more clarity to those issues.


\begin{thebibliography}{10}
\bibitem{key-72}C. Rovelli, ``Loop Quantum Gravity'', Living Reviews
in Relativity Vol. 1 (1998) {[}http://www.livingreviews.org/lrr-1998-1{]}

\bibitem{key-73}R. Arnowitt, S. Deser, C. Misner, \textquotedbl{}Dynamical
Structure and Definition of Energy in General Relativity\textquotedbl{}.
Physical Review 116 (5): 1322\textendash{}1330

\bibitem{key-74}G. Ponzano, T. Regge, ``Semiclassical limit of Racah
coefficients'', Spectroscopic and Group Theoretical Methods in Physics,
pp.1-58

\bibitem{key-75}S. Carlip, ``Lectures in (2+1)-dimensional gravity'',
J.Korean Phys.Soc.28:S447-S467,1995 {[}gr-qc/9503024{]}

\bibitem{key-76}H. Ooguri, ``Topological lattice models in four-dimensions'',
Mod. Phys. Lett. A7, 2799\textendash{}2810 (1992), {[}hep-th/9205090{]}

\bibitem{key-77}L. Crane and D. Yetter, ``A categorical construction
of 4d TQFTs'', Quantum Topology eds. L. Kauff{}- man and R. Baadhio,
World Scientifi{}c, Singapore, 1993, pp. 120\textendash{}130. {[}hep-th/9301062{]}

\bibitem{key-78}J.C. Baez, ``An introduction to spin foam models
of quantum gravity and BF theory'', Lect. Notes Phys., vol 543, p.
25, 2000 {[}gr-qc/9905087{]}

\bibitem{key-79}J. Engle, E. Livine, R. Pereira, C. Rovelli, ``LQG
vertex with finite Immirzi parameter'', Nucl. Phys. vol. B799, pp.
136-149, 2008 {[}gr-qc/0711.0146{]}

\bibitem{key-80}L. Freidel, K. Krasnov, ``A New Spin Foam Model
for 4D Gravity'', Class. Quant. Grav. vol. 25, p. 125018, 2008 {[}gr-qc/0708.1595{]}

\bibitem{key-81}E. Livine, S. Speziale, ``A new spinfoam vertex
for quantum gravity'', Phys. Rev. D vol. 76, p. 084028, 2007 {[}gr-qc/0705.0674{]}

\bibitem{key-82}M. Han, M. Zhang, ``Asymptotics of Spin Foam Amplitude
on Simplicial Manifold: Euclidean Theory'' {[}gr-qc/1109.0500{]}

\bibitem{key-83}M. Han, M. Zhang, ``Asymptotics of Spin Foam Amplitude
on Simplicial Manifold: Lorentzian Theory'' {[}gr-qc/1109.0499{]}

\bibitem{key-84}C. Rovelli, L. Smolin, ``Loop space representation
of quantum general relativity'', Nucl. Phys. B vol. 331, issue 1,
pp. 80-152.

\bibitem{key-85}J.W. Barrett, R. Dowdall, W. Fairbairn, H. Gomes,
F. Hellmann, ``Asymptotic analysis of the EPRL four-simplex amplitude'',
J. Math. Phys. 50:112504, 2009 {[}gr-qc/0902.1170{]}

\bibitem{key-86}K. Giesel, S. Hofmann, T. Thiemann and O. Winkler,
``Manifestly Gauge-Invariant General Relativistic Perturbation Theory'',
{[}arXiv:0711.0115{]}, {[}arXiv:0711.0117{]}

\bibitem{key-87}T. Thiemann, ``Quantum Spin Dynamics'' {[}arXiv:gr-qc/9606089{]}
{[}arXiv:gr-qc/9606090{]}

\bibitem{key-88}L. Kauffman, ``State models and the Jones polynomial'',
Topology 26 (1987), no. 3, pp. 395-407.

\bibitem{key-89}J.W. Barrett, R. Dowdall, W. Fairbairn, F. Hellmann,
R. Pereira, ``Lorentzian spin foam amplitudes: graphical calculus
and asymptotics'' {[}gr-qc/0907.2440{]}

\bibitem{key-90}E. Bianchi, D. Regoli, C. Rovelli, ``Face amplitude
of spinfoam quantum gravity'', Class. Quant. Grav. 27:185009, 2010
{[}gr-qc/1005.0764{]}

\bibitem{key-91}V. Bonzom, ``Spin foam models for quantum gravity
from lattice path integrals'', Phys. Rev. D 80:064028, 2009 {[}gr-qc/0905.1501{]}

\bibitem{key-92}T. Regge, ``General Relativity Without Coordinates'',
Il Nuovo Cimento, Vol. 19, N. 3, p. 558

\bibitem{key-93}F. Conrady, L. Freidel, ``Path integral representation
of spin foam models of 4D gravity'', Class. Quant. Grav. 25, 245010,
2008 {[}gr-qc/0806.4640{]}.

\bibitem{key-94}F. Conrady, L. Freidel, ``On the semiclassical limit
of 4D spin foam models'', Phys. Rev. D 78, 104023, 2008 {[}gr-qc/0809.2280{]}

\bibitem{key-95}J.W. Barrett, L. Crane, ``Relativistic Spin Networks
and Quantum Gravity'', J.Math.Phys. 39, 3296-3302, 1998 {[}gr-qc/9709028{]}

\bibitem{key-96}V. Guillemin, S. Sternberg, ``Symplectic Techniques
in Physics'', Cambridge University Press 1990

\bibitem{key-97}L. Hörmander, ``The Analysis of Linear Partial Differential
Operators I: Distribution Theory and Fourier Analysis'', Springer-Verlag,
2nd Edition.

\bibitem{key-98}G. Lachaud, ``Exponential Sums as Discrete Fourier
Transform with Invariant Phase Functions'', Proceedings of the 10th
International Symposium on Applied Algebra, Algebraic Algorithms and
Error-Correcting Codes, pp. 231-243

\bibitem{key-99}International Journal of Mathematics and Mathematical
Sciences, Volume 2003 (2003), Issue 49, Pages 3091-3099 {[}http://math.iisc.ernet.in/\textasciitilde{}rangaraj/docs/ijmms\_so\_n.pdf{]}

\bibitem{key-100}E. Magliaro, C. Perini, ''Curvature in spinfoams'',
Class.Quant.Grav. 28 (2011) 145028 {[}gr-qc/1103.4602{]}

\bibitem{key-101}F. Hellmann, W. Kaminski, ''Holonomy spin foam
models: Asymptotic geometry of the partition function'' {[}arXiv:1307.1679{]}

\bibitem{key-102}B. Bahr, F. Hellmann, W. Kaminski, M. Kisielowski,
J. Lewandowski, Operator Spin Foam Models {[}arXiv:1010.4787{]}

\bibitem{key-104}L. van Elfrinkhof, Eene eigenschap van de orthogonale
substitutie van de vierde orde. Handelingen van het 6e Nederlandsch
Natuurkundig en Geneeskundig Congres, Delft, 1897. {[}http://www.xs4all.nl/\%7Ejemebius/Elfrinkhof.htm{]}\end{thebibliography}
\end{document}